\documentclass[aps,prd,preprint,groupedaddress,nofootinbib]{revtex4-1}

\usepackage{graphicx}
\usepackage{epsfig}
 \usepackage{bm}
\usepackage{amsfonts}
\usepackage{amsmath}
 \usepackage{subfigure}
\usepackage{wasysym}

\begin{document}



\title{Non-Gaussianity and Secondary Gravitational Waves from Primordial Black Holes Production in $\alpha$-attractor Inflation}





\author{
Kazem Rezazadeh$^{1}$\footnote{kazem.rezazadeh@ipm.ir},
Zeinab Teimoori$^{2}$\footnote{zteimoori16@gmail.com},
Saeid Karimi$^{2}$\footnote{saeedkarimi248@gmail.com},
and Kayoomars Karami$^{2}$\footnote{kkarami@uok.ac.ir}
}

\affiliation{
$^{1}$\small{School of Physics, Institute for Research in Fundamental Sciences (IPM),
P.O. Box 19395-5531, Tehran, Iran}\\
$^{2}$\small{Department of Physics, University of Kurdistan, Pasdaran Street, P.O. Box 66177-15175, Sanandaj, Iran}
}


\date{\today}


\begin{abstract}

We study the non-Gaussianity and secondary Gravitational Waves (GWs) in the process of the Primordial Black Holes (PBHs) production from inflation. In our work, we focus on the $\alpha$-attractor inflation model in which a tiny bump in the inflaton potential enhances the amplitude of the curvature perturbations at some scales and consequently leads to the PBHs production with different mass scales. We implement the computational code BINGO which calculates the non-Gaussianity parameter in different triangle configurations. Our examination implies that in this setup, the non-Gaussianity gets amplified significantly in the equilateral shape around the scales in which the power spectrum of the scalar perturbations undergoes a sharp declination. The imprints of these non-Gaussianities can be probed in the scales corresponding to the BBN and $\mu$-distortion events, or in smaller scales, and detection of such signatures in the future observations may confirm the idea of our model for the generation of PBHs or rule it out. Moreover, we investigate the secondary GWs in this framework and show that in our model, the peak of the present fractional energy density is obtained as $\Omega_{\rm GW0} \sim 10^{-8}$ at different frequencies which depends on the model parameters. These results lie well within the sensitivity region of some GWs detectors at some frequencies, and therefore the observational compatibility of our model can be evaluated by the forthcoming data from these detectors. We further provide some estimations for the tilts of the induced GWs spectrum in the different intervals of frequency, and demonstrate that the spectrum obeys the power-law relation $\Omega_{\rm GW0}\sim f^{n}$ in those frequency bands.

\end{abstract}



\maketitle



\section{Introduction}
\label{section:introduction}

The idea that a considerable part of the Dark Matter (DM) can be constituted from Primordial Black Holes (PBHs) has absorbed much interests in the recent years \cite{Bird:2016dcv, Sasaki:2016jop, Blinnikov:2016bxu, Carr:2016drx, Clesse:2016vqa, Carr:2017jsz, mroz2017no, Garcia-Bellido:2017mdw, Germani:2017bcs, Clesse:2017bsw, Ezquiaga:2017fvi, Gong:2017qlj, ballesteros2018primordial, Zumalacarregui:2017qqd,  Ballesteros:2018wlw, ashoorioon2021eft, Ashoorioon:2020hln, Cai:2018tuh, Kamenshchik:2018sig, Niikura:2019kqi, Chen:2019zza, mishra2020primordial, Ballesteros:2017fsr, Bhaumik:2019tvl, germani2017primordial, garcia2017primordial, Motohashi:2017kbs, dalianis2019primordial, Mahbub:2019uhl, Kawai:2021edk, Motohashi:2019rhu, Liu:2020oqe, Fu:2019ttf, dalianis2020generalized, Dalianis:2019vit, Teimoori:2021thk, Lin:2020goi, Yi:2020cut, Gao:2020tsa, Gao:2021vxb, Solbi:2021rse, Solbi:2021wbo, Teimoori:2021pte, Heydari:2021gea, Heydari:2021qsr}. The first detection of Gravitational Wave (GW), GW150914, from merging of binary black holes with mass $\sim 30 M_\odot$ where $M_\odot$ is the solar mass, by the LIGO-Virgo Collaboration \cite{Abbott:2016blz, Abbott:2016nmj}, has attracted more attention to the physics of PBHs and increased the possibility that the origin of some part of DM or even all of it may be PBHs seeded in the early universe. Besides, the failure of direct detection of particle DM amplifies this idea. PBHs can also be regarded as a DM candidate at other mass scales. Due to the Hawking radiation \cite{Hawking:1975vcx}, light PBHs are expected to be evaporated by now, and this fact together with the implications of the recent observations \cite{gould1992femtolensing, dalcanton1994observational, alcock2001macho, nemiroff2001limits, wilkinson2001limits, tisserand2007limits, carr2010new, griest2013new, Jacobs:2014yca, Ali-Haimoud:2016mbv, Wang:2016ana, Sato-Polito:2019hws, Laha:2020ivk} put tight constraints on the range of mass distribution of dark matter PBHs. PBHs with earth mass ($\sim\mathcal{O}\left(10^{-5}\right)M_{\astrosun}$) can comprise of order $\mathcal{O}\left(10^{-2}\right)$ of DM observed today. Moreover, PBHs can constitute a substantial portion of DM in the two asteroid mass ranges $10^{-16}-10^{-14}M_{\astrosun}$ and $10^{-13}-10^{-11}M_{\astrosun}$ \cite{Niikura:2017zjd, Katz:2018zrn, Laha:2019ssq, Dasgupta:2019cae, Carr:2009jm, Graham:2015apa}.

The origin of PBHs formation may return to the primordial perturbations generated in the inflationary stage of the Universe. The relevant scales to those perturbations exit the Hubble horizon at some epochs during inflation and reenter the horizon later in the radiation-dominated era. The gravitational collapse of those perturbations after their reentry gives rise to PBHs formation if they have sufficiently large amplitude. This idea was first proposed by Zel'dovich and Novikov in 1966, and then by Hawking and Carr in the early 1970s \cite{zel1967hypothesis, Hawking:1971ei, Carr:1974nx}. However, providing such a large amplitude usually raises some difficulties, because CMB observations constrain the amplitude of the scalar perturbations to be of order ${\cal P}_{s}(k_{*})=\mathcal{P}_{s*}\sim\mathcal{O}\left(10^{-9}\right)$ at the pivot scale $k_{*}=0.05\,\mathrm{Mpc}^{-1}$ \cite{akrami2020planck}, while in order to have PBHs, their amplitude should grow seven orders of magnitude and reach $\mathcal{P}_{s}\sim\mathcal{O}\left(10^{-2}\right)$ at some smaller scales.

So far, many suggestions have been proposed to overcome this problem and provide a significant growth in the curvature power spectrum giving rise ultimately to PBHs formation in some scales during inflation. These suggestions are mainly characterized by different considerations for the statistics of the curvature perturbations, because the statistical feature of the primordial perturbations can affect on the abundance of PBHs \cite{Saito:2008em, Byrnes:2012yx, Young:2013oia, Tada:2015noa, Young:2015kda, Young:2015cyn, Garcia-Bellido:2017aan, Franciolini:2018vbk, ezquiaga2018quantum, Passaglia:2018ixg, Atal:2018neu, Belotsky:2018wph, DeLuca:2019qsy, Yoo:2019pma, Ezquiaga:2019ftu, Serpico:2020ehh, Braglia:2020taf, Figueroa:2020jkf, Davies:2021loj, Zhang:2021rqs}. The PBHs formation due to the drastic amplification of the Gaussian power spectrum in the framework of the single-field inflation has been well investigated in \cite{Cai:2018tuh, Kamenshchik:2018sig, Niikura:2019kqi, Chen:2019zza, mishra2020primordial, Ballesteros:2017fsr, Bhaumik:2019tvl, dalianis2020generalized, germani2017primordial, garcia2017primordial, Motohashi:2017kbs, dalianis2019primordial, Mahbub:2019uhl, Kawai:2021edk, Motohashi:2019rhu, Liu:2020oqe, Fu:2019ttf, dalianis2020generalized, Dalianis:2019vit, Teimoori:2021thk, Lin:2020goi, Yi:2020cut, Gao:2020tsa, Gao:2021vxb, Solbi:2021rse, Solbi:2021wbo, Teimoori:2021pte, Heydari:2021gea, Heydari:2021qsr}. In the literature, most studies focus on the scenarios in which inflation takes place in a multi-phase process, and the slow-roll condition is violated at least in one phase so that the inflaton field experiences a so-called ultra slow-roll (or friction-dominated) phase that leads to efficient growth in the primordial power spectrum, see e.g. \cite{germani2017primordial, garcia2017primordial, Motohashi:2017kbs, Bhaumik:2019tvl, Ballesteros:2017fsr, dalianis2019primordial, Mahbub:2019uhl, Kawai:2021edk, Motohashi:2019rhu, Liu:2020oqe, Fu:2019ttf, dalianis2020generalized, Dalianis:2019vit, Teimoori:2021thk, Lin:2020goi, Yi:2020cut, Gao:2020tsa, Gao:2021vxb, Solbi:2021rse, Solbi:2021wbo, Teimoori:2021pte, Heydari:2021gea}. For instance, in nonminimal derivative coupling inflation, the mechanism of the gravitationally enhanced friction leads to an ultra slow-roll inflationary phase which in turn provides a remarkable enhancement of the power spectrum amplitude that is large enough for PBHs production \cite{Fu:2019ttf, dalianis2020generalized, Dalianis:2019vit, Teimoori:2021thk}. The origin of PBHs may come back to the existence of an inflection-point in the inflaton potential \cite{germani2017primordial, garcia2017primordial, Motohashi:2017kbs, Bhaumik:2019tvl, Ballesteros:2017fsr, dalianis2019primordial, Mahbub:2019uhl}. This idea has already been investigated in the framework of $\alpha$-attractor inflation \cite{dalianis2019primordial, Mahbub:2019uhl}. The $\alpha$-attractors belong to the class of supergravity and superconformal inflationary theories which have been proposed based on the conformal symmetry in the Jordan frame, and have a universal attractor behavior in the Einstein frame \cite{Kallosh:2013hoa, Kallosh:2013yoa, Kallosh:2014rga, Kallosh:2014laa, Kallosh:2015lwa, Linde:2015uga, Galante:2014ifa, Carrasco:2015rva, Carrasco:2015pla, Roest:2015qya, scalisi2015cosmological, Eshaghi:2016kne}. The free parameter $\alpha$ is inversely proportional to the curvature of the inflaton K\"{a}hler manifold \cite{Kallosh:2013yoa, Galante:2014ifa}. The $\alpha$-attractor models, in the large $e$-fold number $N$ and small $\alpha$ limit, lead to the same inflationary predictions including the scalar spectral index and tensor-to-scalar ratio as $n_s=1-2/N$ and $r = 12\alpha/N^2$, respectively, which for $\alpha=1$, these inflationary observables coincide exactly with the predictions of the Starobinsky model \cite{Starobinsky:1980te} as well as the Higgs inflation scenario \cite{Bezrukov:2007ep}.

PBHs also can be seeded in the multiple-field inflationary frameworks \cite{Garcia-Bellido:1996mdl, Sasaki:2006kq, Kawasaki:2006zv, Kawaguchi:2007fz, Frampton:2010sw, Lyth:2011kj, Bugaev:2011wy, Kohri:2012yw, Kawasaki:2012wr, Linde:2012bt, Bugaev:2013vba, Clesse:2015wea, Garcia-Bellido:2016dkw, Kawasaki:2016pql, Inomata:2016rbd, Domcke:2017fix, Braglia:2020eai, Ahmed:2021ucx}. In these settings, the sufficiently large peaks required for the PBHs formation can be supplied both with curvature (adiabatic) perturbations \cite{Polarski:1992dq, Starobinsky:1994mh} and isocurvature ones \cite{Starobinsky:1994mh, Kofman:1986wm, Kofman:1988xg, Polarski:1994rz, Starobinsky:2001xq}. In particular, in the context of the more traditional hybrid inflation, the authors of \cite{Clesse:2015wea} have succeeded within a two-field scenario to prepare a large peak in the scalar power spectrum. It should be noted that the multi-field inflationary models are capable to produce strong non-Gaussianities in the curvature power spectrum, and the curvaton scenario is a well-motivated example for these class of models \cite{Young:2013oia, Sasaki:2006kq}.

Most of the scenarios of PBHs formation from inflation are based on the assumption of Gaussian statistics for the curvature perturbations. However, the spectrum of the curvature fluctuations may follow from the non-Gaussian statistics \cite{Saito:2008em, Byrnes:2012yx, Young:2013oia, Tada:2015noa, Young:2015kda, Young:2015cyn, Garcia-Bellido:2017aan, Franciolini:2018vbk, ezquiaga2018quantum, Passaglia:2018ixg, Atal:2018neu, Belotsky:2018wph, DeLuca:2019qsy, Yoo:2019pma, Ezquiaga:2019ftu, Serpico:2020ehh, Braglia:2020taf, Figueroa:2020jkf, Davies:2021loj, Zhang:2021rqs}. Recently, the authors of \cite{Garcia-Bellido:2017aan} considered the effect of primordial non-Gaussianity on the PBHs abundance and concluded that the non-Gaussian spectrum can provide a similar abundance for PBHs from a lower amplitude of the scalar perturbations relative to the Gaussian spectrum. It is crucial to point out that due to the impact of non-Gaussianity on the PBHs abundance, the formation of PBHs can be regarded as an instrument to evaluate the amplitude and non-Gaussianity of the curvature perturbations at the small scales.

The production of PBHs due to enhancement of the power spectrum of the curvature perturbations during inflation may be accompanied by induction of secondary GWs \cite{Matarrese:1997ay, Mollerach:2003nq, Ananda:2006af, Baumann:2007zm, Saito:2008jc, Saito:2009jt, Bugaev:2009zh, Bugaev:2010bb, Alabidi:2012ex, Nakama:2016gzw, Peirone:2017vcq, Cheng:2018yyr}. Consequently, if we observe PBHs or secondary GWs, then it is possible to use them to constrain the amplification of the curvature perturbations power spectrum during inflation which provides valuable information about the physics of the early universe for us. Various ideas are suggested so far to generate secondary GWs by virtue of the production of PBHs from inflation. One idea is application of a phenomenological delta function giving rise to the required growth in the primordial power spectrum \cite{Saito:2008jc, Bartolo:2018evs, Cai:2018dig}. This goal can also be achieved appropriately by using a broken power-law \cite{Lu:2019sti} or Gaussian power spectrum \cite{Namba:2015gja, Garcia-Bellido:2017aan, Lu:2019sti}. The running-mass model \cite{Stewart:1996ey, Drees:2011hb, Datta:2019euh}, the axion-curvaton model \cite{Kasuya:2009up, Kawasaki:2012wr}, oscillons after inflation \cite{Easther:2006vd, Antusch:2016con, Liu:2017hua}, and parametric resonance due to the double inflation \cite{Kawasaki:2006zv, Kawasaki:2016pql} are another mechanisms leading to the generation of PBHs and secondary GWs in the context of inflationary cosmology. Note that, like the abundance of PBHs, the spectrum of induced GWs depends on both the amplitude and statistics of the scalar quantum fluctuations \cite{Garcia-Bellido:2017aan}. However, in contrast to the PBHs abundance which is very sensitive to the statistics of the curvature perturbations, the amplitude of the secondary GWs is not very sensitive to this feature and it is hardly affected by the non-Gaussianity of the primordial curvature perturbations \cite{Zhang:2021vak}.

It is possible to produce PBHs only with a trivial modification in the potential shape, like the approach of \cite{mishra2020primordial} in which the source of DM originates from a tiny bump in the potential of the $\alpha$-attractor inflation. In this scenario, the bump behaves like a speed-breaker and slows down the scalar field evolution. It leads to a sharp increase in the amplitude of the curvature perturbations that is required for PBHs production.

In the present work, we are interested to calculate the primordial non-Gaussianity on the different scales and also the secondary gravitational waves within the setup of the $\alpha$-attractor inflation model in the presence of a tiny bump in its potential. The main reason for us to concentrate this model originates from the fact that this model firstly has well-based theoretical motivation from string theory \cite{Kallosh:2013hoa, Kallosh:2013yoa}. Secondly, this model is in excellent agreement with the latest CMB data released from the Planck 2018 \cite{akrami2020planck} measurements. The prediction of this model for the scalar spectral index $n_s$ and the tensor-to-scalar ratio $r$, is located satisfactorily within the 68\% CL marginalized joint region of the Planck 2018 data \cite{akrami2020planck}, as it has been shown in \cite{mishra2020primordial}. Most of the inflationary models predicts rather large values for $r$ so that they are ruled out explicitly by the current observations. The $\alpha$-attractor scenario gives a small value for this observable, and therefore it remains in the class of acceptable inflationary models in light of the current data. This model is further capable to provide smaller values for $r$, and so if the future data prefer less value for this observable, the $\alpha$-attractor model can still be remain compatible with the observations.

Although the production of PBHs from $\alpha$-attractor inflation has already been investigated in \cite{mishra2020primordial}, there exist some aspects of this model that has not been regarded so far. Moreover, although the authors of \cite{mishra2020primordial} have computed the curvature power spectrum $\mathcal{P}_s$ of this model, they did not compare their findings with the current upper bounds on this spectrum. At the CMB scales, this spectrum is tightly constrained by the Planck 2018 observations \cite{akrami2020planck}. At the smaller scales, there are still a lot of constraints on $\mathcal{P}_s$, such as the bounds from CMB spectral distortions \cite{Kohri:2014lza,Chluba:2012we}, Acoustic Reheating (AR) \cite{Inomata:2016uip}, European Pulsar Timing Array (EPTA) observations \cite{lentati2015european}, Big Bang Nucleosynthesis (BBN) \cite{Kohri:2018awv}, and advanced Laser Interferometer Gravitational wave Observatory (aLIGO) O2 \cite{LIGOScientific:2017zlf,LIGOScientific:2017vwq}. In addition, there are several sensitivity regions for the curvature perturbation spectrum from different detectors like Square Kilometre Array (SKA) \cite{Moore:2014lga,Janssen:2014dka}, the Big Bang Observer (BBO) \cite{Yagi:2011wg,phinney2003big}, The DECi-hertz Interferometer Gravitational wave Observatory (DECIGO) \cite{Seto:2001qf,Yagi:2011wg}, Einstein Telescope (ET) \cite{Punturo:2010zz}, Laser Interferometer Space Antenna (LISA) \cite{LISA:2017pwj}, and aLIGO design  \cite{KAGRA:2013rdx}, that can be used to check viability of PBHs scenarios when their data are released. Especially, the sensitivity of ET on the scales of $10^{14}-10^{19} \mathrm{Mpc}^{-1}$ is high enough to evaluate the amplitude of $\sim 10^{-5}$ for the spectrum of the curvature perturbations. Therefore, it will be much useful if the result of our model for the scalar power spectrum is compared with these sensitivity regions.

Besides, it will be useful if the results of the PBHs abundances $f_{\rm{PBH}}$ in different mass scales are compared with the current constraints on $f_{\rm{PBH}}$ such as the constraints of Extra-Galactic gamma-ray (EG$\gamma$) \cite{Carr:2009jm}, White Dwarf explosion (WD) \cite{Graham:2015apa}, microlensing events with Subaru Hyper Suprime-Cam (HSC) \cite{Niikura:2017zjd}, Kepler milli/microlensing (Kepler) \cite{Griest:2013esa}, the Earth Resources Observation System(EROS)/Massive Astrophysical Compact Halo Object (MACHO) microlensing (EROS) \cite{EROS-2:2006ryy}, the Optical Gravitational Lensing Experiment (OGLE) \cite{Niikura:2019kqi}, dynamical heating of Ultra-Faint Dwarf galaxies (UFD) \cite{Brandt:2016aco}, and accretion constraints by CMB \cite{Ali-Haimoud:2016mbv, aloni2017cosmic, Poulin:2017bwe}. The results of the model also can be appraised in light of the aLIGO O1 \& O2 \cite{LIGOScientific:2016dsl, LIGOScientific:2017zlf, LIGOScientific:2017vwq}, aLIGO design \cite{KAGRA:2013rdx}, ET and Cosmic Explorer (CE) \cite{LIGOScientific:2016wof} detectable limits. Accordingly, in the present work, we examine the possibility of PBHs production with masses around ${\cal O}(10)M_\odot$, ${\cal O}(10^{-5})M_\odot$, and ${\cal O}(10^{-13})M_\odot$ in the framework of the $\alpha$-attractor inflation with a tiny bump in the potential, in light of the current constraints on $\mathcal{P}_s$ and $f_{\rm{PBH}}$. We show that for the model satisfy the current constraints on the curvature power spectrum and the PBHs abundance, the model parameters should be somewhat different from those used in \cite{mishra2020primordial}.

Furthermore, we study the subjects of non-Gaussianity and secondary GWs in the process of PBHs formation in the framework of $\alpha$-attractor Inflation. These concepts are very important issues which have absorbed much attentions recently (see, e.g. \cite{Saito:2008em, Byrnes:2012yx, Young:2013oia, Tada:2015noa, Young:2015kda, Young:2015cyn, Garcia-Bellido:2017aan, Franciolini:2018vbk, ezquiaga2018quantum, Passaglia:2018ixg, Atal:2018neu, Belotsky:2018wph, DeLuca:2019qsy, Yoo:2019pma, Ezquiaga:2019ftu, Serpico:2020ehh, Braglia:2020taf, Figueroa:2020jkf, Davies:2021loj, Zhang:2021rqs}). The non-Gaussianity parameter is an important observable that can provide valuable information about the inflationary era of the early Universe. We also investigate the secondary GWs in this framework. We calculate the present fractional energy density of these GWs and compare our findings with the sensitivity regions of different GWs detectors. Additionally, we estimate the slope of the secondary GWs spectrum in different frequency ranges. This feature of the secondary GWs spectrum also may be assessed by the forthcoming data, and then can be used to discriminate between the different mechanisms of PBHs production from inflation. In the recent years, several GWs detectors have been designed, and their data can be used to scrutinize the PBHs seeded from the primordial perturbations. Among these detectors, one can refer to aLIGO O2 \& O3 runs \cite{LIGOScientific:2017zlf,LIGOScientific:2017vwq,KAGRA:2021kbb}, the HLV (LIGO-Hanford, LIGOLivingston and Virgo), the mid-scale upgrade of aLIGO (A+ detectors), ET \cite{Punturo:2010zz}, LISA \cite{LISA:2017pwj}, EPTA \cite{lentati2015european},  North American Nanohertz Observatory for Gravitational Waves (NANOGrav) \cite{NANOGrav:2020gpb}, International Pulsar Timing Array (IPTA), and the Square Kilometer Array (SKA) \cite{Moore:2014lga,Janssen:2014dka,Schmitz:2020syl}. Our results in this paper for the abundance of PBHs and the induced GWs spectrum can be produced by our numerical codes which are available online\footnote{\url{https://github.com/krezazadeh/PBHs_alpha_attractor}}.

To compute the non-Gaussianity parameter in various cosmological scales, we employ the numerical code BINGO which is developed by Hazra et al. \cite{Hazra:2012yn} and is publicly available\footnote{\url{https://github.com/dkhaz/bingo}}. This code computes the inflationary non-Gaussianities in different triangle configurations. On the scales far from the bump, the slow-roll regime is completely valid, and hence the non-Gaussianity is small as expected for the single-field standard inflation model \cite{Maldacena:2002vr, Acquaviva:2002ud, Rigopoulos:2004ba}. But on the scales around the bump, the slow-roll condition may be disturbed, and therefore the non-Gaussianity may take large values. The values of non-Gaussianity on these scales may be extensively comparable in front of the values evaluated on the CMB scales ($k_{*}\sim 0.05\,\mathrm{Mpc}^{-1}$), which are strongly constrained by the observational data of Planck satellite \cite{Planck:2019kim}. However, the prediction of large non-Gaussianity on the scales beyond the CMB scales may be checked with more precise measurements in the future. In this way, one can discriminate between the inflationary models proposed to explain PBHs production. Besides, such measurement provides a useful insight for us to understand whether the PBHs can be regarded as the origin of DM or not.

The rest of this paper is structured as follows: first, in Sec. \ref{sec:PBHAbundance}, we present a brief review of the base formulas for the calculation of the mass and abundance of PBHs. In Sec. \ref{section:singlefeildinflation}, we review briefly single field inflationary model. In Sec. \ref{sec:PBHalpha}, we discuss the process of PBH production in the framework of the $\alpha$-attractor scenario with a tiny bump in the potential. Then, in Sec. \ref{section:non-Gaussianity}, we calculate the non-Gaussianity parameter for three different cases of the model which have different values for the parameters. We subsequently investigate the secondary GWs in our model in Sec \ref{sec:sgws}. Finally, in Sec. \ref{section:conclusions}, we summarize our concluding remarks.


\section{Abundance of Primordial Black Holes}
\label{sec:PBHAbundance}

If the amplitude of the scalar modes of perturbations produced during the phase of inflation is large enough, they may gravitationally collapse to form PBHs after horizon re-entry in the radiation dominated era. The relation between the mass of formed PBHs, $M(k)$ and the horizon mass, $M_{\rm hor}$ is \cite{Gong:2017qlj}
\begin{equation}
\label{PBHmass}
M(k)=\gamma M_{\rm hor}=3.68\left(\frac{\gamma}{0.2}\right)\left(\frac{g_*}{10.75}\right)^{-1/6}\left(\frac{k}{10^{6}\,\,\rm{Mpc^{-1}}}\right)^{-2}M_{\odot},
\end{equation}
where $\gamma$ is the efficiency factor and $g_*$ is the effective degrees of freedom for the energy density. These parameters usually assumed to be $\gamma \simeq 0.2 $ and $g_*=107.5$ \cite{carr1975primordial}.

We assume that the curvature perturbations obey Gaussian statistics. In this way the production rate of PBHs with mass $M(k)$ has the following form \cite{Young:2014ana, ozsoy2018mechanisms, Tada:2019amh}
\begin{equation}\label{beta}
\beta(M)=\sqrt{\frac{2}{\pi}} \frac{\sigma(M)}{\delta_c}\exp{\left(-\frac{\delta_c^{2}}{2\sigma^2(M)}\right)},
\end{equation}
where $\delta_c$ is the threshold of the density perturbation for the PBHs formation, and we assume to be $\delta_c=0.4$ \cite{Musco:2012au, Harada:2013epa, Escriva:2019phb}. $\sigma^2(k)$ is the coarse-grained variance of the density contrast smoothed on a scale $k$ and is given by \cite{Young:2014ana, ozsoy2018mechanisms}
\begin{equation}\label{sigma2}
\sigma^2(k)=\int{\frac{dq}{q}}\,W^2(q/k)\frac{16}{81}(q/k)^4 {\cal P}_s(q),
\end{equation}
where ${\cal P}_s$ is the power spectrum of the curvature perturbations and $W(x)$ is the window function. In this work we choose Gaussian window function as $W(x)=\exp{(-x^2/2)}$ that is very popular in the literature. Other choices for window functions have been studied in \cite{Young:2019osy,Ando:2018qdb}.

The current energy fraction of PBHs with a mass $M(k)$ over the total DM is given by \cite{Gong:2017qlj, Carr:2016drx}
\begin{equation}\label{fPBH}
f_{\rm{PBH}}(M)\equiv \frac{\Omega_{\rm {PBH}}}{\Omega_{\rm{DM}}}=\frac{\beta(M)}{3.94\times10^{-9}}\left(\frac{\gamma}{0.2}\right)^{1/2}\left(\frac{g_*}{10.75}\right)^{-1/4}\left(\frac{0.12}{\Omega_{\rm{DM}}h^2}\right)
\left(\frac{M}{M_{\odot}}\right)^{-1/2},
\end{equation}
where $\Omega_{\rm {DM}}$ is the current density parameter of DM and we take its value as $\Omega_{\rm {DM}}h^2\simeq0.12$ from the Planck 2018 results \cite{akrami2020planck}.


\section{Single Field Inflationary Model}
\label{section:singlefeildinflation}

The standard single field inflationary paradigm is described by the action \cite{Baumann:2009ds}
\begin{equation}\label{action}
S= \int {\rm d}^{4}x \sqrt{-g}\left[\frac{1}{2} R + X-V(\phi) \right],
\end{equation}
where $g$ is the determinant of the metric $ g_{{\mu}{\nu}}$, $R$ is the Ricci scalar, $X\equiv -\frac{1}{2}g^{\mu\nu}\partial_\mu{\phi}\partial_\nu{\phi}$ is the kinetic term, and $V(\phi)$ is the scalar field potential. Throughout this paper, we set the reduced Planck mass equal to unity, i.e., $M_{\rm{pl}}=(8\pi G)^{-1/2}=1$.

For the flat Friedmann-Robertson-Walker (FRW) metric $g_{{\mu}{\nu}}={\rm diag}\Big(-1, a^{2}(t), a^{2}(t), a^{2}(t)\Big)$, we have $X=\dot{\phi}^2/2$ and the background equations turn into \cite{Baumann:2009ds}
\begin{align}
\label{eq:FR1}
 & 3 H^{2}-\frac{1}{2}\dot{\phi}^{2}-V(\phi) =0, \\
  \label{eq:FR2}
 & 2 \dot{H}+3H^2+\frac{1}{2}\dot{\phi}^{2}-V(\phi) =0,\\
  \label{eq:Field}
 &\ddot{\phi}+3H\dot{\phi}+V_{,\phi}=0,
\end{align}
where $H\equiv \dot{a}/a $ is the Hubble parameter and the overdot represents the derivative with respect to the cosmic time $t$.

In the context of standard single field inflation, the slow-roll parameters are defined as \cite{Baumann:2009ds}
\begin{equation}\label{SRparameters}
\varepsilon_{H} \equiv -\frac{\dot H}{H^2}, \hspace{.5cm}  \eta_{H}\equiv -\frac{\ddot{\phi}}{ H\, \dot{\phi}}.
\end{equation}
The first relation in Eq. \eqref{SRparameters} indicates that to have inflation, the condition $\varepsilon<1$ should be satisfied. Note that, during the slow-roll regime, two parameters $\varepsilon_{H}$ and $\eta_{H}$ should be much smaller than unity. Using Eqs. \eqref{eq:FR1} and \eqref{eq:FR2}, one can easily find the slow-roll parameter $\varepsilon _{H}$ as
\begin{equation}\label{epsilon:exact}
\varepsilon_{H} =\frac{\dot{\phi}^2}{2H^2}.
\end{equation}
In the framework of standard inflation, the power spectrum of the scalar perturbation ${\cal P}_{s}$, is in the following form \cite{Baumann:2009ds}
\begin{equation}\label{Ps1}
{\cal P}_{s}=\frac{H^2}{8 \pi ^{2}}\frac{1}{\varepsilon_{H}}\Big|_{k=aH},
\end{equation}
where ${\cal P}_{s}$ is computed at the time of sound horizon exit, i.e., $k = aH$ in which $k$ denotes the comoving wavenumber. The observational value of the amplitude of scalar perturbations at the CMB pivot scale $k_{*}=0.05\,{\rm Mpc}^{{\rm -1}}$ reported by the Planck team is ${\cal P}_{s}(k_{*})\simeq 2.1 \times 10^{-9}$ \cite{akrami2020planck}.

Under the slow-roll approximation, the field equations \eqref{eq:FR1} and \eqref{eq:Field} reduce to
\begin{align}
\label{FR1:SR}
& 3 H^2\simeq V(\phi),\\
  \label{Field:SR}
& 3 H\dot{\phi}+V_{,\phi}\simeq0.
\end{align}
Using Eqs. \eqref{FR1:SR} and \eqref{Field:SR}, the slow-roll parameter $\varepsilon_{H}$ in (\ref{epsilon:exact}) can be written as
\begin{equation}\label{epsilonv}
\varepsilon_{H} \simeq \varepsilon_{V}\equiv \frac{1}{2}\left(\frac{V_{,\phi}}{V}\right)^2.
\end{equation}
With the help of the first Friedmann Eq. \eqref{FR1:SR} and Eq. \eqref{epsilonv}, the power spectrum \eqref{Ps1} in the slow-roll limit reduces to
\begin{equation}\label{PsSR}
{\cal P}_{s}\simeq \frac{V^3}{12\pi^2 V_{,\phi}^2}.
\end{equation}
In the slow-roll approximation, the Hubble parameter $H$ and its variation change much slower than the scale factor $a$ of the Universe \cite{Garriga:1999vw}. Therefore, using the relation $k=a H$, we have $d\ln k\approx H dt$. Applying this approximation and the definition $n_s-1 \equiv d\ln{\cal P}_{s}/d\ln k$, one can easily find $n_s-1\simeq \dot{{\cal P}_{s}}/(H{\cal P}_{s})$. Using this relation and Eqs. \eqref{FR1:SR}, \eqref{Field:SR}, \eqref{PsSR}, and the definition \eqref{epsilonv}, we can find the scalar spectral index $n_s$ as
\begin{equation}\label{nsSR}
n_s-1\simeq 2\eta_{V}-6\varepsilon_{V},
\end{equation}
where
\begin{equation}\label{etav}
\eta_{V}\equiv \frac{V_{,\phi\phi}}{V}=\eta_{H}-\varepsilon_{H}.
\end{equation}
The observational value of the scalar spectral index is $n_{s}=0.9627\pm0.0060$ (68\% CL, Planck 2018 TT+lowE) \cite{akrami2020planck}. Using the approximation $d\ln k\approx H dt$ and Eq. \eqref{Field:SR}, the running of the scalar spectral index as
\begin{equation}\label{alphas}
\frac{d{n_s}}{d\ln k}\simeq -\left(\frac{V_{,\phi}}{V(\phi)}\right) n_{s,\phi}.
\end{equation}
The observational constraint on the running of the scalar spectral index is about $d{n_s}/d \ln k= - {\rm{0}}{\rm{.0078}} \pm 0.0082$ (68\% CL, Planck 2018 TT+lowE) \cite{akrami2020planck}.

The amount of inflation is usually described by the $e$-fold number $N$ which is defined as
\begin{equation}\label{efold-definition}
N \equiv \ln \left(\frac{a_e}{a}\right),
\end{equation}
where $a_{e}$ is the scale factor at the end of inflation. The definition \ref{efold-definition} is equivalent to
\begin{equation}\label{efold}
dN=-H\,dt.
\end{equation}
It is important to note that the number of $e$-folds between the horizon exit and the end of inflation should be about $50-60$ \cite{Dodelson:2003vq, Liddle:2003as}.

The tensor power spectrum in the framework of the standard canonical inflationary model is given by
\begin{equation}\label{Pt}
{\cal P}_{t}=\frac{H^2}{2\pi ^{2}}\Big|_{k=aH}.
\end{equation}
From Eqs. \eqref{Ps1} and \eqref{Pt}, we get the tensor-to-scalar ratio in the slow-roll regime as
\begin{equation}\label{r}
r = 16 \varepsilon_{H} \simeq 16\varepsilon_V.
\end{equation}
The Planck 2018 data determined an upper bound on the tensor-to-scalar ratio as $r< 0.0654$ (68\% CL, Planck 2018 TT+lowE) \cite{akrami2020planck}.


\section{$\alpha$-attractor inflation with a tiny bump in the potential}
\label{sec:PBHalpha}

Including a tiny bump, the general form of the potential can be written as the following form \cite{mishra2020primordial}
\begin{equation}
 \label{V}
 V(\phi)=V_{b}(\phi)\left[1+\varepsilon(\phi)\right],
\end{equation}
where $V_b$ is the base potential of the model and here it is taken in the form of $\alpha$-attractor potential \cite{Kallosh:2013hoa, Kallosh:2013yoa}
\begin{equation}
 \label{Vb}
 V_{b}(\phi)=V_{0}\tanh^{2n}\left(\frac{\phi}{\sqrt{6\alpha}}\right).
\end{equation}
This is the original inflationary potential responsible for providing primordial quantum fluctuations compatible with the CMB constraints on the inflationary observables including the scalar spectral index $n_s$ and tensor-to-scalar ratio $r$. Following \cite{mishra2020primordial}, we set two free parameters $\alpha$ and $n$ to be $\alpha = 1$ and $n=1$. The function $\varepsilon$ in \eqref{V} specifies the shape of the bump, and it is considered as \cite{mishra2020primordial}
\begin{equation}
 \label{varepsilon}
 \varepsilon(\phi)=A\cosh^{-2}\left(\frac{\phi-\phi_{0}}{\sigma}\right),
\end{equation}
where $A$, $\sigma$, and $\phi_{0}$ characterize the height, width, and position of the peak, respectively. As discussed in \cite{mishra2020primordial}, the presence of such a modification in the potential function, leads to sufficiently large enhancement in the curvature power spectrum required for the PBHs formation.

In \cite{mishra2020primordial}, the authors have considered three cases for the model that have different values for these parameters, and estimated the PBHs abundance for each case. In the present paper, we take the amounts of these parameters as listed in Table \ref{table:parameters}. The value of $V_0$ is fixed by imposing the CMB normalization at the pivot scale $k_{*}=0.05\,{\rm Mpc}^{{\rm -1}}$ corresponding to $N_*=60$. It is worth to mentioning that the values for Case 1 and Case 3 are very close to those adopted in \cite{mishra2020primordial}, but we have changed the values slightly here in order to the cases satisfy the existing constraints on the curvature power spectrum as well as the bounds on the PBHs abundances at different cosmological scales. In addition, we have considered another parameter set as Case 2 whose parameters are completely different with those adopted in \cite{mishra2020primordial}. We will show this case provides properly the current observational bounds too. We further compute the inflationary observables $n_s$, $r$, and $dn_s/d\ln k$ at the $k_{*}=0.05\,{\rm Mpc}^{{\rm -1}}$, and also quantities relevant for producing PBHs of the considered three cases. The results are presented in Table \ref{Table:nsrfM}.

\begin{table}[ht!]
  \centering
  \caption{The successful parameter sets for the potential \eqref{V} that can produce PBHs. The value of $V_0$ is fixed by imposing the CMB normalization at the pivot scale $k_{*}=0.05\,{\rm Mpc}^{{\rm -1}}$ corresponding to $N_*=60$.}
\scalebox{1}[1] {
    \begin{tabular}{ccccc}
    \hline
    \hline
    $\qquad \# \qquad$ & $\qquad \phi_{0} \qquad$ & $\qquad \qquad A \qquad \qquad$ & $\qquad \qquad \sigma \qquad \qquad$ & $\qquad V_{0}\qquad$\tabularnewline
    \hline
 Case 1 & $4.850001$ & $2.044880\times10^{-3}$ & $2.524999\times10^{-2}$ & $1.448\times10^{-10}$\tabularnewline
    \hline
 Case 2 & $5.301327$ & $1.080552\times10^{-3}$ & $1.940550\times10^{-2}$ & $1.495\times10^{-10}$\tabularnewline
    \hline
  Case 3 & $5.580000$ & $6.402075\times10^{-4}$ & $1.428997\times10^{-2}$ & $1.420\times10^{-10}$\tabularnewline
    \hline
    \end{tabular}
    }
  \label{table:parameters}
\end{table}

\begin{figure*}
\begin{minipage}[b]{1\textwidth}
\subfigure{\includegraphics[width=.48\textwidth]%
{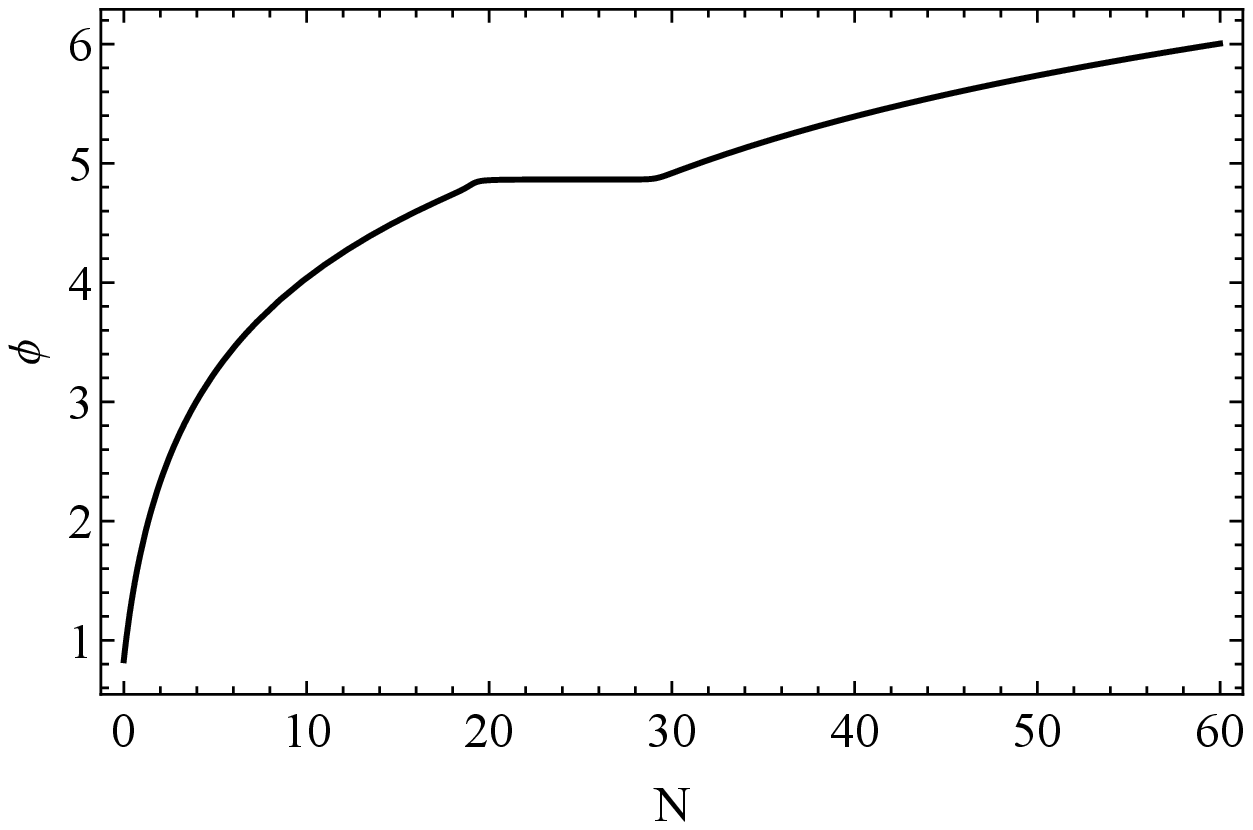}} \hspace{.1cm}
\subfigure{\includegraphics[width=.48\textwidth]%
{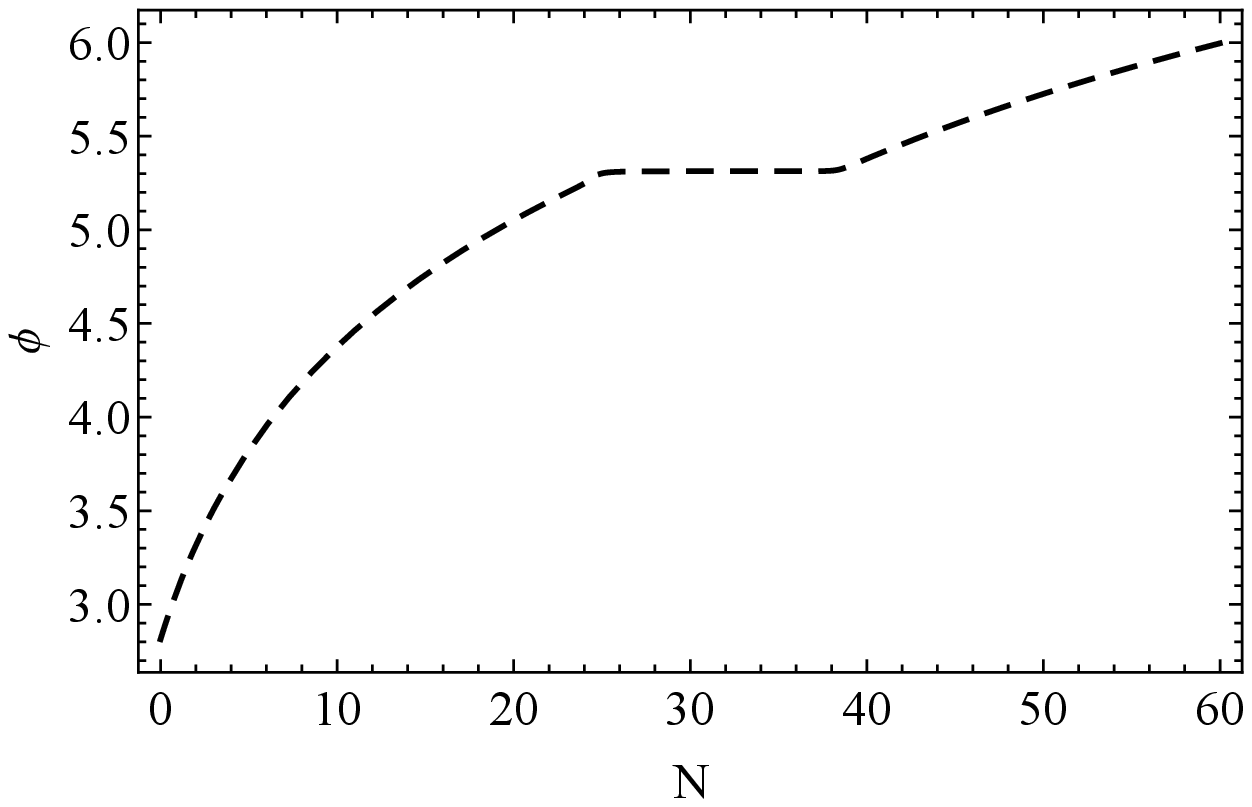}}\hspace{.1cm}
\subfigure{\includegraphics[width=.48\textwidth]%
{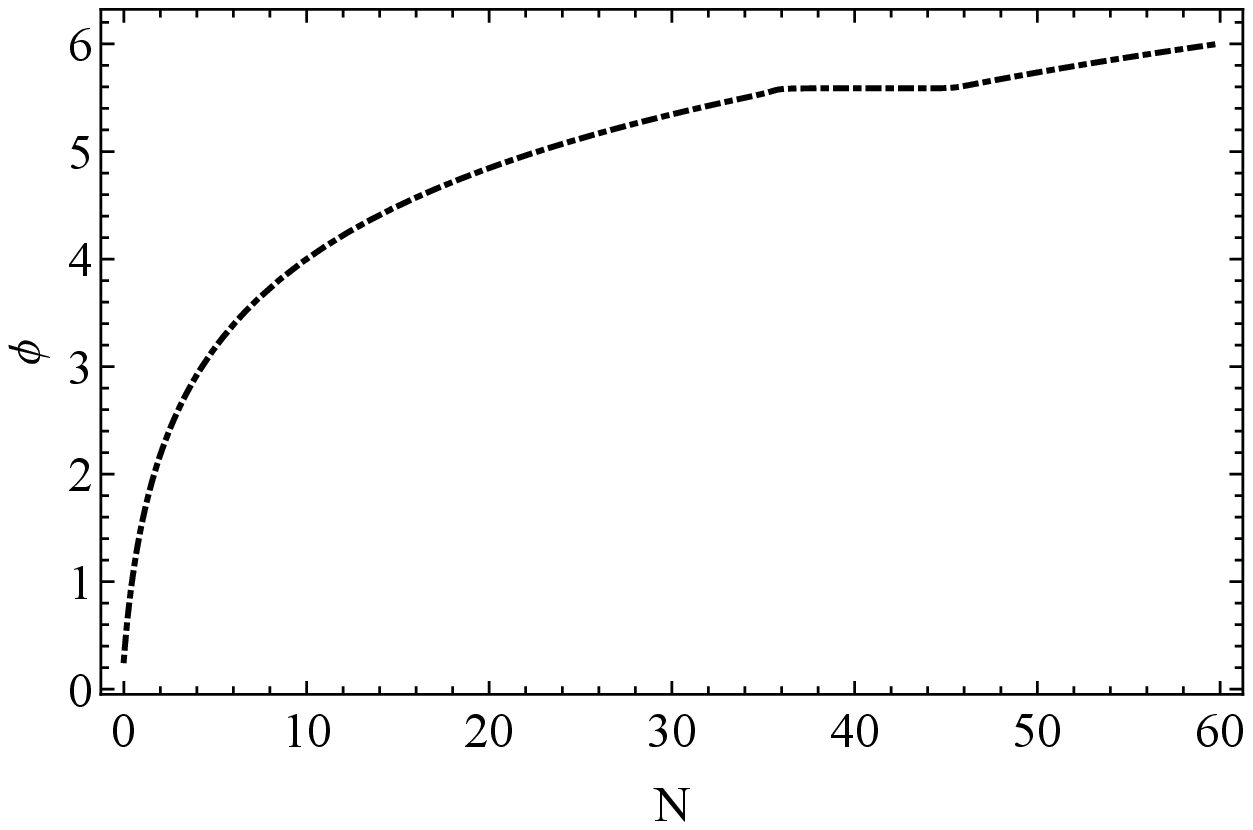}}
\end{minipage}
\caption{Evolution of the scalar field $\phi$ as a function of the $e$-fold number $N$ for Case 1 (solid line), Case 2 (dashed line), and Case 3 (dash-dotted line).}
\label{fig:phiN}
\end{figure*}

The evolution of the scalar field $\phi$ as a function of the $e$-fold number $N$, for Case 1 (solid line), Case 2 (dashed line), and Case 3 (dash-dotted line) are plotted in Fig. \ref{fig:phiN} by solving the full equations of motion \eqref{eq:FR1}-\eqref{eq:Field} numerically. We set the initial conditions by using Eqs. \eqref{FR1:SR} and \eqref{Field:SR} at $N_{*}=60$. In this figure we see that a plateau-like region appears at $\phi=\phi_0$ corresponding to $20 \lesssim N\lesssim 29$ for Case 1, $26 \lesssim N\lesssim 39$ for Case 2, and $37\lesssim N\lesssim 46$ for Case 3. During this region, the inflaton undergoes a period of ultra slow-roll inflation in which its evolution is slowing down by increasing friction which leads to enhance the curvature power spectrum by several orders of magnitude that is required to PBHs formation.

\begin{table*}[ht!]
  \centering
  \caption{Results of the scalar spectral index $n_s$, the tensor-to-scalar ratio $r$, the running of the scalar spectral index $dn_s/d\ln k$, the mass of the corresponding PBHs $M_{\rm{PBH}}^{\rm{peak}}$, and the PBH fractional abundance $f_{\rm{PBH}}$  for the three cases of Table \ref{table:parameters}. The inflationary observables $n_s$, $r$, and $dn_s/d\ln k$ are found at the CMB scale corresponding to $N_*=60$.}
\scalebox{1}[1] {\begin{tabular}{c c c c c c}
    \hline
    \hline
     \#  & \,\,\,\,\,\,\,\,\,$n_s$ &\,\,\,\,\,\,\,\,\,  $r$  &\,\,\,\,\,\,\,\,\,  $dn_s/d\ln k$   &\,\,\,\,\,\,\,\,\, $M_{\rm{PBH}}^{\rm{peak}}/M_\odot$ &\,\,\,\,\,\,\,\,\, $f_{\rm{PBH}}^{\rm{peak}}$\\
    \hline
    Case 1 &\,\,\,\,\,\,\,\,\, $0.96031$ &\,\,\,\,\,\,\,\,\, $0.00465$& \,\,\,\,\,\,\,\,\, $-0.000783$ &\,\,\,\,\,\,\,\,\,$1.66\times10^{-13}$ &\,\,\,\,\,\,\,\,\,$0.9550$\\
    Case 2 &\,\,\,\,\,\,\,\,\, $0.96003$ &\,\,\,\,\,\,\,\,\, $0.00471$&\,\,\,\,\,\,\,\,\,  $-0.000794$   &\,\,\,\,\,\,\,\,\,$1.28\times10^{-5}$ &\,\,\,\,\,\,\,\,\, $0.0217$\\
   Case 3 &\,\,\,\,\,\,\,\,\, $0.96028$ &\,\,\,\,\,\,\,\,\, $0.00465$& \,\,\,\,\,\,\,\,\,  $-0.000784$  &\,\,\,\,\,\,\,\,\,$20.98$ &\,\,\,\,\,\,\,\,\,$0.0070$\\
    \hline
    \end{tabular}}
  \label{Table:nsrfM}
\end{table*}

\begin{figure*}[t]
\begin{center}
\scalebox{1}[1]{\includegraphics{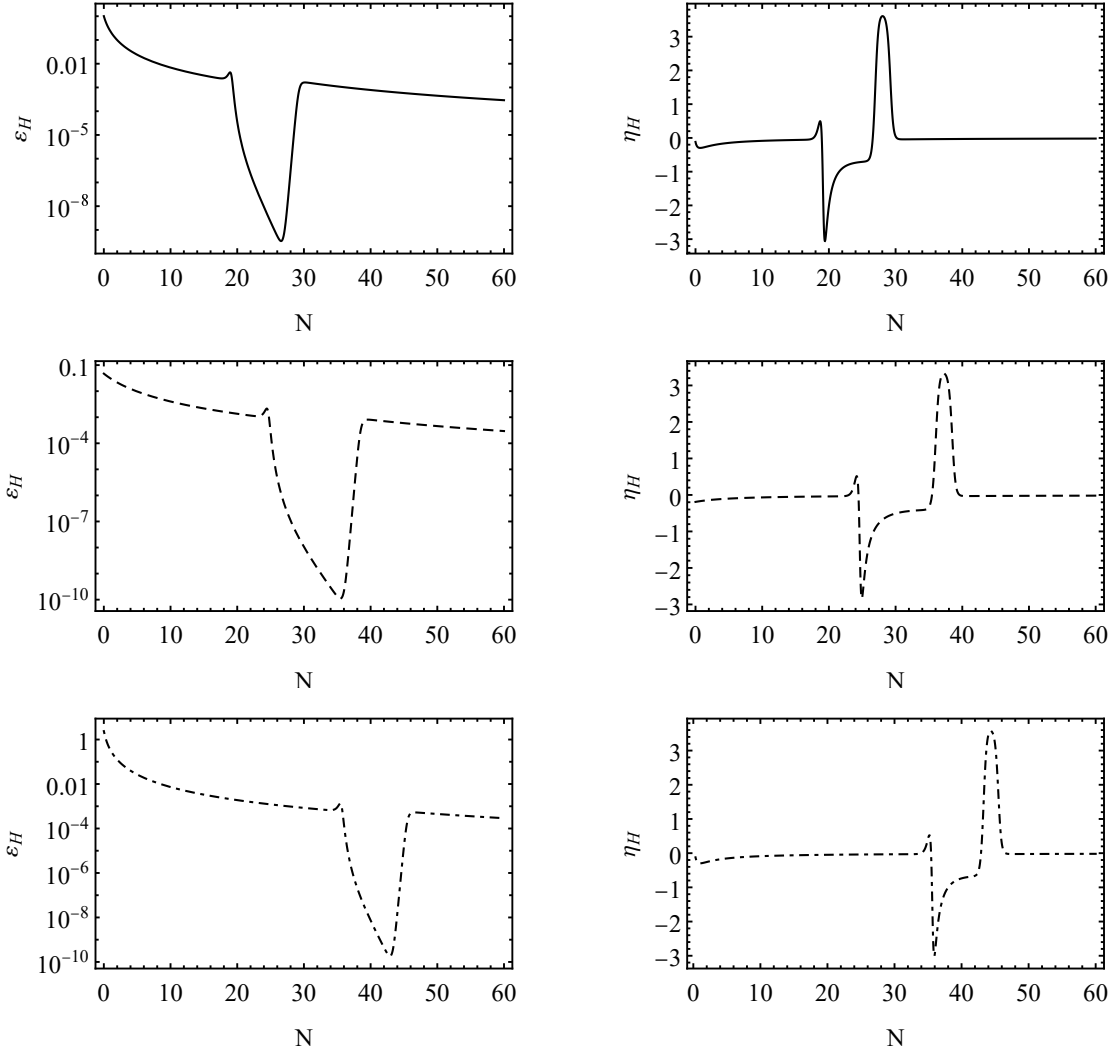}}
\caption{Evolution of the first slow-roll parameter $\varepsilon_{H}$ (left) and the second slow-roll parameter $\eta_{H}$ (right) versus the $e$-fold number $N$ for Case 1 (solid line), Case 2 (dashed line), and Case 3 (dash-dotted line).}
\label{fig:SRparameters}
\end{center}
\end{figure*}

In Fig. \ref{fig:SRparameters} we plot the evolution of the slow-roll parameters $\varepsilon_{H}$ (left panel) and $\eta_{H}$ (right panel) as functions of $N$ for the three cases of Table \ref{table:parameters}. From the left panel of this figure we see that $\varepsilon_{H}$ remains below unity up until the end of inflation $(\varepsilon_{H}<1)$ but the right panel shows that the slow-roll approximation  $\eta_{H}\ll 1$ is violated on the scales of the bump. Of course, it is worth mentioning that for Case 2 the first slow-roll parameter doesn't reach unity at the end of inflation, i.e., $N=0$ and inflation ends after a finite duration. It continues $3.307$ $e$-folds due to strong slow down of inflaton in the ultra slow-roll phase.

A quick look on Fig. \ref{fig:SRparameters} demonstrates that at the time of sound horizon exit corresponding to $N_{*}=60$, the slow-roll approximation remains valid. Therefore, we can compute the values of scalar spectral index $n_s$, the tensor-to-scalar ratio $r$, and the running of the scalar spectral index $dn_s/d\ln k$ with the help of Eqs. \eqref{nsSR}, \eqref{alphas}, \eqref{r}, and also using Eqs. \eqref{epsilonv}, \eqref{etav} for the three cases of Table \ref{table:parameters}. The numerical results are presented in Table \ref{Table:nsrfM}. We see that the scalar spectral index, the tensor-to-scalar ratio, and the running of the scalar spectral index for all three cases are $n_s \approx 0.96$, $r \approx 0.0047$, and $dn_s/d\ln k \approx -0.0008$, respectively. These values are in agreement with the $68\%$ CL constraints of Planck 2018 TT+lowE data \cite{akrami2020planck}.

\begin{figure*}[t]
\begin{center}
\scalebox{0.8}[0.8]{\includegraphics{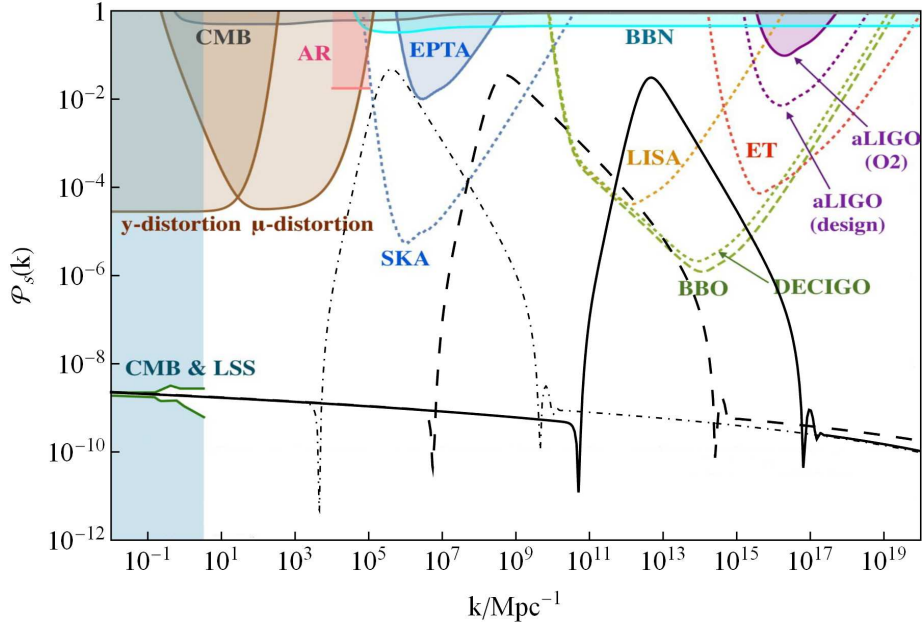}}
\caption{The curvature power spectra founded by solving numerically the Mukhanov-Sasaki equation versus the comoving wavenumber $k$ for Case 1 (solid black line), Case 2 (dashed black line), and Case 3 (dash-dotted black line). The colored shaded regions are excluded by the current observations and the colored dashed lines show the expected limits from the future experiments \cite{Inomata:2018epa}.}
\label{fig:Ps}
\end{center}
\end{figure*}

As discussed above, the slow-roll approximation in this model is violated at the scales of the bump. Therefore, in order to examine the evolution of the curvature perturbation, it is essential to solve numerically the Mukhanov-Sasaki equation which is in the following form
\begin{equation}
 \label{d2vkdtau2}
 v_{k}''+\left(k^{2}-\frac{z''}{z}\right)v_{k}=0,
\end{equation}
where the prime indicates the derivative with respect to the conformal time $\tau$. Also, $v$ is the canonically normalized variable for the curvature perturbation $\mathcal{R}$, and it is defined as $v\equiv z\mathcal{R}$, with $z=a\dot{\phi}/H$. To solve the differential equation \eqref{d2vkdtau2}, we use the Bunch-Daveis vacuum initial condition,
\begin{equation}
 \label{vk-BD}
 v_{k}\to\frac{1}{\sqrt{2k}}e^{-ik\tau},
\end{equation}
which is quite valid for the scales deep inside the horizon. The solution of the Mukhanov-Sasaki equation \eqref{d2vkdtau2} is usually used to calculate the power spectrum of the scalar perturbation which is given by
\begin{equation}
 \label{Ps}
 \mathcal{P}_{s}=\frac{k^{3}}{2\pi^{2}}\frac{\left|v_{k}\right|^{2}}{z^{2}}.
\end{equation}
We evaluate the scalar power spectrum \eqref{Ps} for the three cases of Table \ref{table:parameters}, and plot their scalar power spectrum versus comoving wavenumber $k$ in Fig. \ref{fig:Ps}. The results are shown by the solid (Case 1), dashed (Case 2), and dash-dotted (Case 3) black lines. Note that through this paper, we assume that the pivot scale $k_{*}\sim0.05\,\mathrm{Mpc}^{-1}$ exits the Hubble horizon at $N_* = 60$ $e$-folds before the end of inflation. As we see in Fig. \ref{fig:Ps}, the scalar power spectra of Cases 1, 2, and 3 represent their peaks at the scales $\mathcal{O}(10^{12})$, $\mathcal{O}(10^{8})$, and $\mathcal{O}(10^{5})$ $\mathrm{Mpc}^{-1}$, respectively. In these scales the amplitude of the power spectra get amplified considerably and reach the magnitudes of order $10^{-2}$, while on the CMB scales ($k_{*}\sim0.05\,\mathrm{Mpc}^{-1}$), they approach the value $\mathcal{P}_{s*}\sim\mathcal{O}\left(10^{-9}\right)$ which is compatible with the Planck 2018 measurements \cite{akrami2020planck}. The numerical results for the peak of the scalar power spectrum are presented in Table \ref{table:kPs}. 

In Fig. \ref{fig:Ps}, the existing and expected limits on the curvature power spectrum at the small scales are also specified. The colored shaded regions are excluded by the current observations, whereas the colored dashed lines show the expected limits from future experiments \cite{Inomata:2018epa}. The dark green, brown, pink, blue, light green and purple shaded regions indicate the constraints from the Cosmic Microwave Background/Large-Scale Structure (CMB/LSS) observations \cite{hunt2015search}, CMB spectral distortions \cite{Kohri:2014lza,Chluba:2012we}, Acoustic Reheating (AR) \cite{Inomata:2016uip}, European Pulsar Timing Array (EPTA) observations \cite{lentati2015european}, Big Bang Nucleosynthesis (BBN) \cite{Kohri:2018awv}, and advanced Laser Interferometer Gravitational wave Observatory (aLIGO) O2 \cite{LIGOScientific:2017zlf,LIGOScientific:2017vwq}, respectively. The sensitivity curves of the Square Kilometre Array (SKA) \cite{Moore:2014lga,Janssen:2014dka}, the Big Bang Observer (BBO) \cite{Yagi:2011wg,phinney2003big}, The DECi-hertz Interferometer Gravitational wave Observatory (DECIGO) \cite{Seto:2001qf,Yagi:2011wg}, Einstein Telescope (ET) \cite{Punturo:2010zz}, Laser Interferometer Space Antenna (LISA) \cite{LISA:2017pwj}, and aLIGO design \cite{KAGRA:2013rdx} are also shown by colored dashed lines. We see in the figure that the power spectra of all the three cases satisfy properly the existing constraints, and hence our setup can successfully explain the enhancement of the scalar perturbations required for the PBHs formation. From the figure, we further see that the predictions of our setup are located inside the sensitivity region of SKA, LISA, BBO and DECIGO detectors. Therefore, the viability of our model can be assessed in light of the future observations.

In Fig. \ref{fig:Ps}, we also see that, before the enhancement, the scalar power spectrum shows a sudden declination in its shape, and this is a special feature coming from the existence of the tiny bump in the potential of the $\alpha$-attractor model. We will discuss in the next section that this feature is the one that sources a considerable enhancement for the equilateral non-Gaussianity in this model.

\begin{table}[t]
  \centering
  \caption{The comoving wavenumber and amplitude of the peak of the scalar power spectrum for Case 1, 2, and 3.}
  \scalebox{1.0}{
\begin{tabular}{ccc}
\hline
\#  & $\qquad\qquad\qquad$ $k^{\rm{peak}}\,\left[\mathrm{Mpc}^{-1}\right]$ $\qquad\qquad\qquad$ &  $\mathcal {P}_{s}^{\rm{peak}}$ \tabularnewline
\hline
\hline
Case 1 & $4.833\times10^{12}$ & $3.079\times10^{-2}$ \tabularnewline
\hline
Case 2 & $5.254\times10^{8}$ & $3.518\times10^{-2}$\tabularnewline
\hline
Case 3 & $3.886\times10^{5}$ & $4.740\times10^{-2}$\tabularnewline
\hline
\end{tabular}
  }
  \label{table:kPs}
\end{table}

Using the exact power spectrum obtained by solving the Mukhanov-Sasaki equation numerically and with the help of Eqs. \eqref{PBHmass}-\eqref{fPBH}, we find the mass of the formed PBHs and their abundance for parameter sets 1, 2, and 3. The numerical results are tabulated in Table \ref{Table:nsrfM} which are demonstrated that the model can generate PBHs in a wide range of mass scales. As we see, in Case 1, our model predicts PBHs with the mass $M\simeq 1.66\times10^{-13} M_\odot$ and PBH abundance $f_{\rm{PBH}}\simeq 0.9550$ which means that the produced PBHs in this class constitute around $0.96\%$  of DM content of the present Universe. It is an interesting result and implies that these PBHs can be taken as a suitable candidate for DM.

\begin{figure}[t]
\begin{center}
\scalebox{0.9}[0.9]{\includegraphics{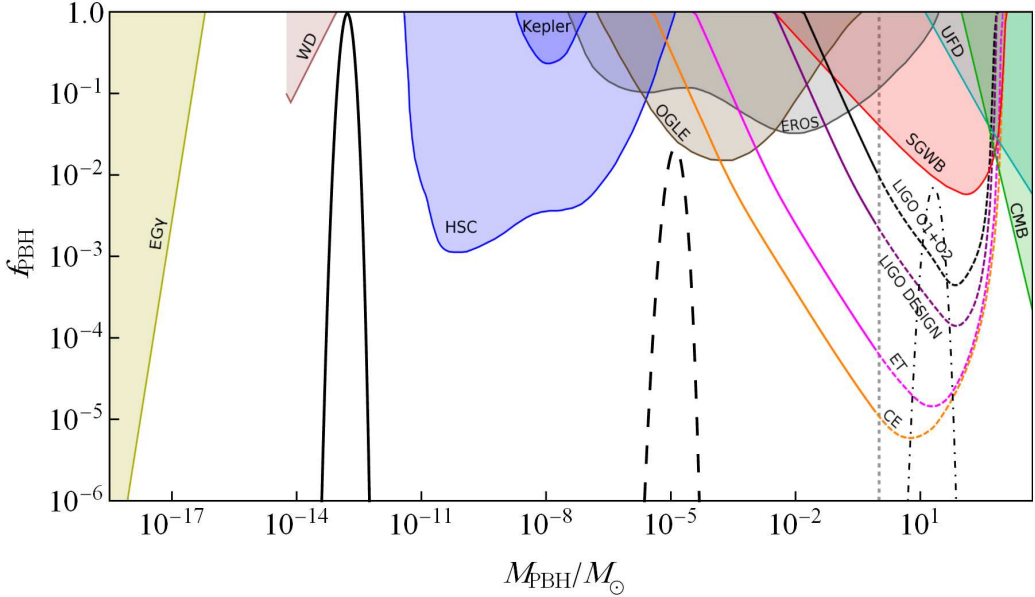}}
\caption{The fractional abundance of PBHs as a function of PBH mass for Case 1 (solid black line), Case 2 (dashed black line), and Case 3 (dash-dotted black line). The black, purple, magenta, and orange curves indicate the results of targeted search from aLIGO O1 \& O2, aLIGO design, ET \cite{Punturo:2010zz}, and CE \cite{LIGOScientific:2016wof}, respectively \cite{Chen:2019irf}. The dashed lines illustrate the detectable limits of $f_{\rm{PBH}}$ from these detectors. The gray vertical line shows that the constraints from the targeted search are only valid for the sub-solar mass PBHs \cite{Chen:2019irf}. The red curve is the updated upper bound of $f_{\rm{PBH}}$ constrained by the non-detection of SGWB from both aLIGO O1 \& O2 searches. The other shaded regions indicate the current observational constraints on the abundance of PBHs: Extra-Galactic gamma-ray (EG$\gamma$) \cite{Carr:2009jm}, White Dwarf explosion (WD) \cite{Graham:2015apa}, microlensing events with Subaru Hyper Suprime-Cam (HSC) \cite{Niikura:2017zjd}, Kepler milli/microlensing (Kepler) \cite{Griest:2013esa}, the Earth Resources Observation System/Massive Astrophysical Compact Halo Object (EROS/MACHO) microlensing (EROS) \cite{EROS-2:2006ryy}, the Optical Gravitational Lensing Experiment (OGLE) \cite{Niikura:2019kqi}, dynamical heating of Ultra-Faint Dwarf galaxies (UFD)\cite{Brandt:2016aco}, and accretion constraints by CMB \cite{Ali-Haimoud:2016mbv, aloni2017cosmic, Poulin:2017bwe}.}
\label{fPBH-M}
\end{center}
\end{figure}

In Fig. \ref{fPBH-M}, we plot the abundance of formed PBHs, $f_{\rm{PBH}}$  versus the mass for Case 1 (solid black line), Case 2 (dashed black line) and Case 3 (dash-dotted black line). In this figure, the current constraints on $f_{\rm{PBH}}$ are also shown \cite{Chen:2019irf}. The black, purple, magenta, and orange curves indicate the results of targeted search from aLIGO O1 \& O2 \cite{LIGOScientific:2016dsl, LIGOScientific:2017zlf, LIGOScientific:2017vwq}, aLIGO design \cite{KAGRA:2013rdx}, ET and Cosmic Explorer (CE) \cite{LIGOScientific:2016wof}, respectively \cite{Chen:2019irf}. The dashed lines illustrate the detectable limits of $f_{\rm{PBH}}$ from these detectors. The gray vertical line shows that the constraints from the targeted search are only valid for the sub-solar mass PBHs \cite{Chen:2019irf}. The red curve is the updated upper bound of $f_{\rm{PBH}}$ constrained by the non-detection of Stochastic Gravitational Wave Background (SGWB) from both aLIGO O1 \& O2 searches \cite{Chen:2019irf}. The other shaded regions indicate the current observational constraints on the abundance of PBHs: Extra-Galactic gamma-ray (EG$\gamma$) \cite{Carr:2009jm}, White Dwarf explosion (WD) \cite{Graham:2015apa}, microlensing events with Subaru Hyper Suprime-Cam (HSC) \cite{Niikura:2017zjd}, Kepler milli/microlensing (Kepler) \cite{Griest:2013esa}, the Earth Resources Observation System(EROS)/Massive Astrophysical Compact Halo Object (MACHO) microlensing (EROS) \cite{EROS-2:2006ryy}, the Optical Gravitational Lensing Experiment (OGLE) \cite{Niikura:2019kqi}, dynamical heating of Ultra-Faint Dwarf galaxies (UFD)\cite{Brandt:2016aco}, and accretion constraints by CMB \cite{Ali-Haimoud:2016mbv,aloni2017cosmic,Poulin:2017bwe}. From the figure, we see that the predictions of our model for the three cases are in agreement with the existing constraints. Moreover, the figure shows that the obtained results for Case 3 can be located within the detectable limits of $f_{\rm{PBH}}$ by the targeted search from aLIGO O1 \& O2, aLIGO design, ET and CE.

\section{Calculation of non-Gaussianity}
\label{section:non-Gaussianity}

Planck 2018 collaboration \cite{Planck:2019kim} has analyzed the cosmic microwave background temperature and $E$-mode polarization maps to obtain constraints on primordial non-Gaussianity. Their combined temperature and polarization analysis produces the following final results for the local, equilateral, and orthogonal bispectrum amplitudes: $f_{\mathrm{NL}}^{\mathrm{local}}=-0.9\pm5.1$, $f_{\mathrm{NL}}^{\mathrm{equil}}=-26\pm47$, and $f_{\mathrm{NL}}^{\mathrm{ortho}}=-38\pm24$ (68\% CL) \cite{Planck:2019kim}. Here, we are interested to calculate the non-Gaussianity parameter for the $\alpha$-attractor inflation model with a tiny bump in its potential, and compare our results at the CMB scales with the Planck 2018 constraints \cite{Planck:2019kim}. To do so, we use the publicly available computational code BINGO \cite{Hazra:2012yn} which is capable to compute the non-Gaussianity parameter in different triangle configurations.

\begin{figure}
\centering
\includegraphics[scale=0.8]{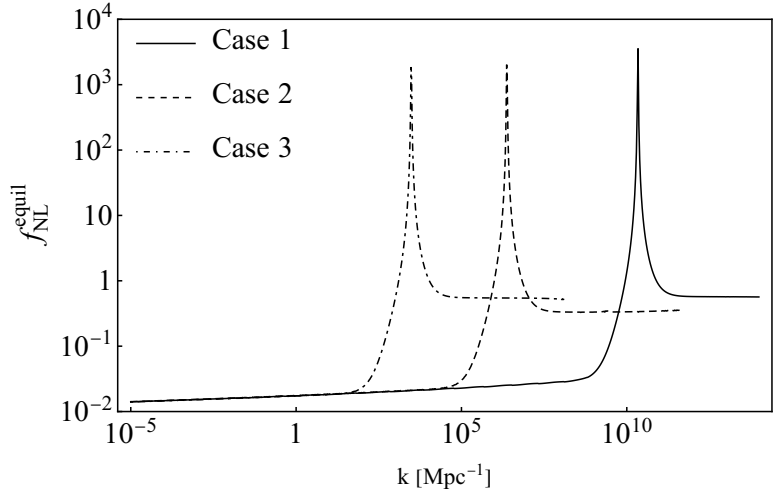}
\caption{The equilateral non-Gaussianity parameter $f_{\mathrm{NL}}^{\mathrm{equil}}$ for the three Cases 1, 2, and 3.}
\label{fig:equilateral}
\end{figure}

First, we use the code to calculate the equilateral non-Gaussianity $f_{\mathrm{NL}}^{\mathrm{equil}}$ for the three cases regarded in this paper. The results of these cases are presented in Fig. \ref{fig:equilateral}. We see in the figure that at the CMB scales, the equilateral non-Gaussianity for the three cases take the values which are of order the slow-roll parameters, $f_{\mathrm{NL}}^{\mathrm{equil}}\sim\mathcal{O}\left(10^{-2}\right)$, and therefore are in excellent agreement with the 68\% CL constraint of Planck 2018 data \cite{Planck:2019kim}. At the smaller scales, however, as we see in the figure, the equilateral non-Gaussianity in each case represents a peak in its diagram which is much larger than unity. For all the three cases the peak of equilateral non-Gaussianity is of order $f_{\mathrm{NL}}^{\mathrm{equil}}\sim\mathcal{O}\left(10^{3}\right)$, but the comoving wavenumber of the peak is different for each case. For Cases 1, 2, and 3, the comoving wavenumber of the peak is equal to $\mathcal{O}(10^{11})$, $\mathcal{O}(10^{10})$, and $\mathcal{O}(10^{4})$ $\mathrm{Mpc}^{-1}$, respectively. Our numerical findings are reported in Table \ref{table:fNLequil}. If we compare these values with the wavenumbers of the peaks in Table \ref{table:kPs}, we see that the wavenumbers at which the equilateral non-Gaussianity peaks are close to the one at which the scalar power spectrum peaks. More precisely, the peaks of the equilateral non-Gaussianity appear at the scales where the power spectrum of the scalar perturbations undergoes a sharp declination in its shape before it enhances and reaches its maximum value. Our conclusion about the large equilateral non-Gaussianities in the production of PBHs from the models with a sudden declination is in agreement with the result of \cite{Davies:2021loj} for an ultra-slow-roll inflation model, and with the findings of \cite{Zhang:2021rqs} in the setup of Gauss-Bonnet inflation. The existence of such a sharp declination appears also in some other models like constant-roll \cite{Motohashi:2019rhu} and ultra constant-roll \cite{Liu:2020oqe} inflations. We expect that in these models the equilateral non-Gaussianity presents a substantial enhancement too, that can be examined in future investigations. But there are some models, such as the non-minimal derivative \cite{Fu:2019ttf, dalianis2020generalized, Dalianis:2019vit, Teimoori:2021thk} and $k/G$ \cite{Lin:2020goi, Yi:2020cut, Gao:2020tsa, Gao:2021vxb, Solbi:2021rse, Solbi:2021wbo, Teimoori:2021pte, Heydari:2021gea, Heydari:2021qsr} inflationary scenarios, in which the variation of the scalar power spectrum is rather mild over all scales, and so we expect that these models do not show a large enhancement in the equilateral non-Gaussianity and it remains of order $f_{\mathrm{NL}}^{\mathrm{equil}}\lesssim1$ in all scales. This point also may be verified by further examinations in the future.

\begin{table}[t]
  \centering
  \caption{The comoving wavenumber and amount of non-Gaussianity parameter for the peak of the equilateral non-Gaussianity in Case 1, 2, and 3.}
  \scalebox{1.0}{
\begin{tabular}{lcc}
\hline
\# & $\qquad\qquad\qquad$  $k\,\left[\mathrm{Mpc}^{-1}\right]$ $\qquad\qquad\qquad$ &   $f_{\mathrm{NL}}^{\mathrm{equil}}$\tabularnewline
\hline
\hline
Case 1 &  $2.190\times10^{10}$ &   $3577.85$ \tabularnewline
\hline
Case 2 &  $2.330\times10^{6}$ &  $2022.86$\tabularnewline
\hline
Case 3 &  $2.999\times10^{3}$ &  $1864.40$\tabularnewline
\hline
\end{tabular}
  }
  \label{table:fNLequil}
\end{table}

\begin{figure}
\centering
\includegraphics[scale=0.6]{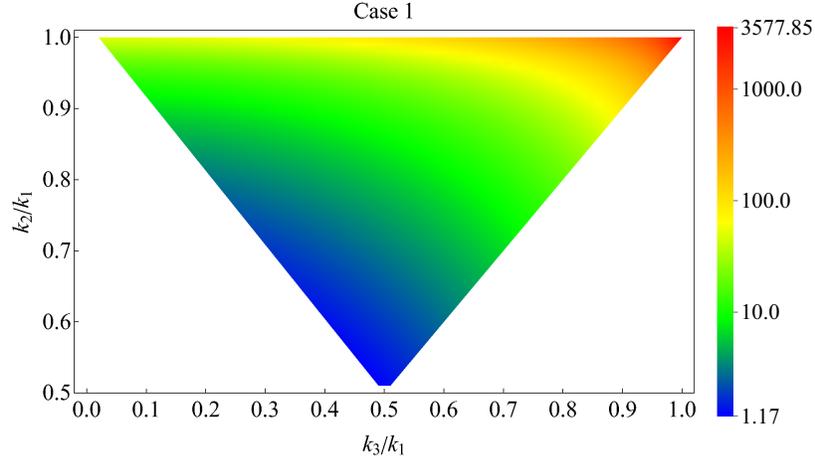}
\caption{The non-Gaussianity in the general form of scalene triangle configuration for Case 1.}
\label{fig:scalene-case1}
\end{figure}

\begin{figure}
\centering
\includegraphics[scale=0.6]{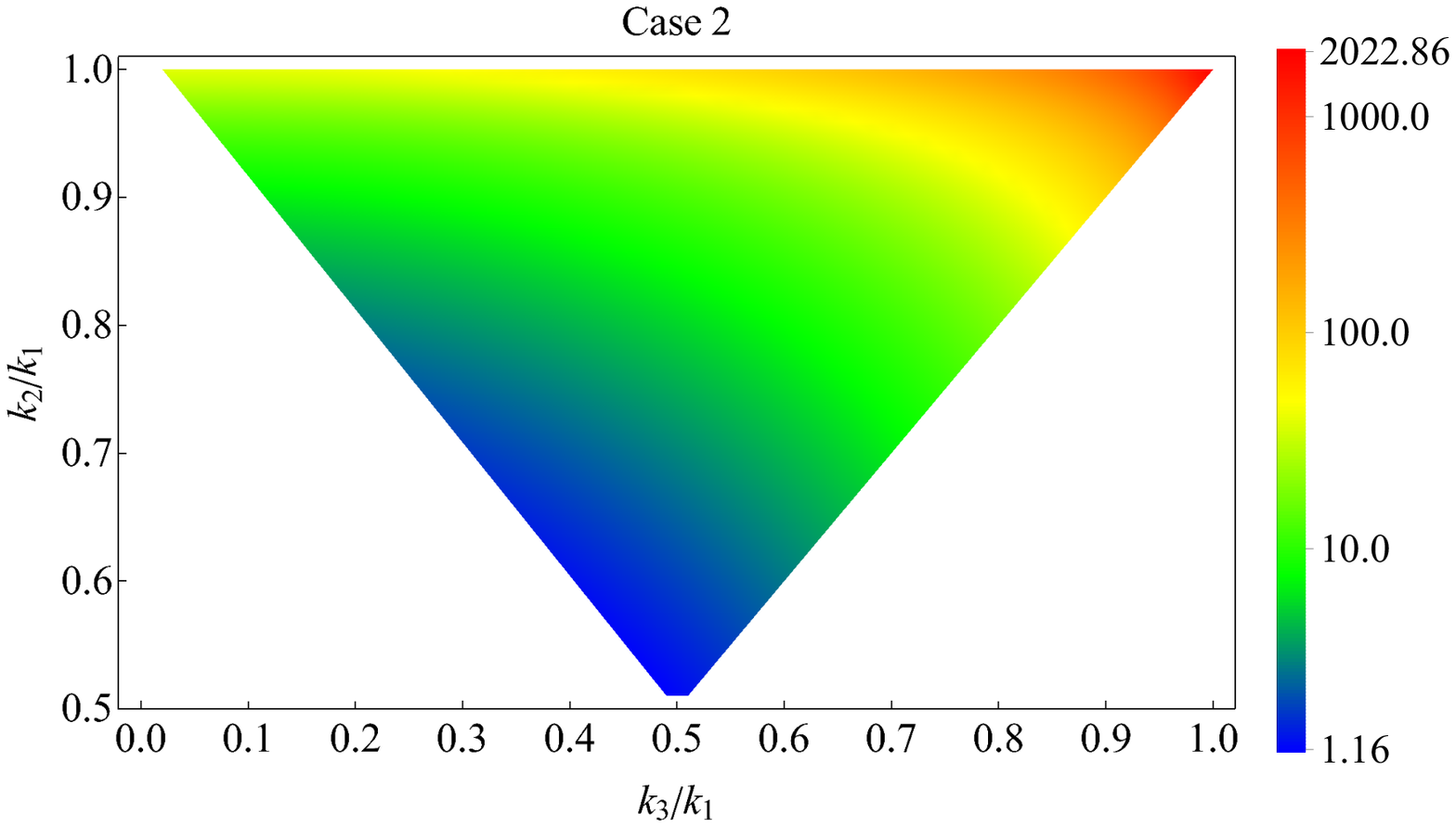}
\caption{The non-Gaussianity in the general form of scalene triangle configuration for Case 2.}
\label{fig:scalene-case2}
\end{figure}

\begin{figure}
\centering
\includegraphics[scale=0.6]{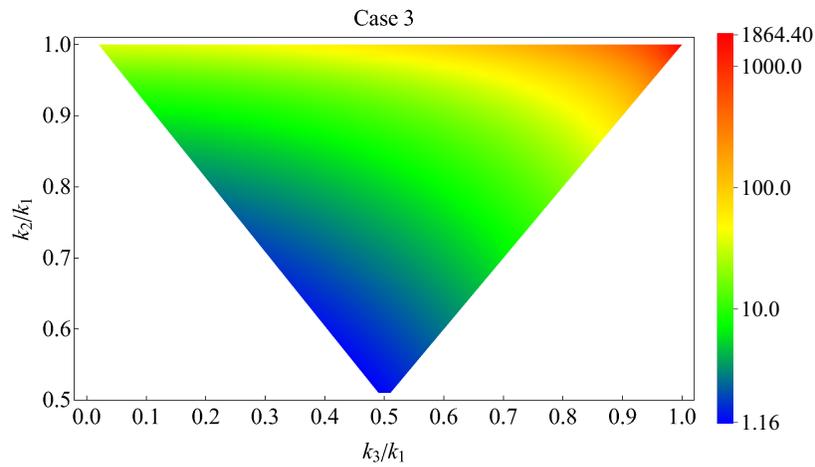}
\caption{The non-Gaussianity in the general form of scalene triangle configuration for Case 3.}
\label{fig:scalene-case3}
\end{figure}

In the next step, we turn to compute the non-Gaussianity in the general form of scalene triangle configuration in which the triangle sides $k_1$, $k_2$, and $k_3$ can take different values that satisfy the triangular inequalities. The parameter $k_1$ in these figures is fixed to the comoving wavenumber at which the equilateral non-Gaussianity peaks and its numerical values are presented in Table \ref{table:fNLequil}. The results of Cases 1, 2, and 3 in this configuration are presented in Figs. \ref{fig:scalene-case1}, \ref{fig:scalene-case2}, and \ref{fig:scalene-case3}, respectively. The top-left point in these figures is related to the squeezed configuration ($k_{1}\approx k_{2}\gg k_{3}$), and the signal of the local non-Gaussianity $f_{\mathrm{NL}}^{\mathrm{local}}$ is maximized in this configuration \cite{Babich:2004gb, Bartolo:2004if, Baumann:2009ds}. In other words, if we see any enhancement at that point in comparison to the other points of the diagram, then the local non-Gaussianity is the most important form in that model. The squeezed configuration is the dominant mode in the multi-field inflationary models \cite{Bartolo:2001cw, Sasaki:2008uc}, the curvaton scenario \cite{Linde:1996gt, Lyth:2002my}, the models with inhomogeneous reheating \cite{Dvali:2003em, Kofman:2003nx}, and the New Ekpyrotic frameworks \cite{Creminelli:2007aq, Koyama:2007if, Buchbinder:2007at}.

The top-right corner in these diagrams corresponds to the equilateral shape ($k_{1}=k_{2}=k_{3}$) \cite{Babich:2004gb, Bartolo:2004if, Baumann:2009ds}, and this kind of non-Gaussianity arises in the models with higher-derivative interactions and non-trivial sound speeds \cite{Alishahiha:2004eh, Chen:2006nt, Li:2008qc, Tolley:2009fg, Amani:2018ueu, Rasouli:2018kvy}. The bottom middle-point in the diagrams are related to the folded triangle form ($k_{1}=2k_{2}=2k_{3}$), and a signal at this point for a model implies that the orthogonal non-Gaussianity $f_{\mathrm{NL}}^{\mathrm{ortho}}$ is the dominant kind in that model \cite{Baumann:2009ds}. This type of non-Gaussianity finds importance in the inflation models with non-standard initial states \cite{Chen:2006nt, Holman:2007na}. Additionally, there are the intermediate cases, namely the elongated ($k_{1}=k_{2}+k_{3}$) and isosceles ($k_{1}>k_{2}=k_{3}$) that their signatures are pronounced respectively along with the left and right sides of the triangle in the figures. In our setting for PBHs production, since the bispectrum enhances at the top-right point, as we see in Figs. \ref{fig:scalene-case1}, \ref{fig:scalene-case2}, and \ref{fig:scalene-case3}, the equilateral non-Gaussianity is the dominant kind.

From Fig. \ref{fig:equilateral}, or also from Figs. \ref{fig:scalene-case1}, \ref{fig:scalene-case2}, and \ref{fig:scalene-case3}, we see that the value of the equilateral non-Gaussianity in our model can reach very large values. This enhancement appears at the scales related to BBN \cite{Jeong:2014gna, Inomata:2016uip}, $\mu$-distortion \cite{Fixsen:1996nj}, or smaller scales. Around the CMB scales, the non-Gaussianity is very small and so it is in very good agreement with the Planck 2018 constraints \cite{Planck:2019kim}. Therefore, we conclude that in order for our model can explain the formation of PBHs, then we can observe large equilateral non-Gaussianities in some special scales. In other words, the prediction of large equilateral non-Gaussianities in those scales is one important new feature of our model. This is a key feature to check the viability of the model in light of future observations. If such non-Gaussianities are observed at the BBN \cite{Jeong:2014gna, Inomata:2016uip} and $\mu$-distortion \cite{Fixsen:1996nj} scales or smaller scales in future experiments, then our model will be confirmed or will be constrained. Conversely, if we cannot detect such a signal in the forthcoming data, the model will be ruled out. In addition, detection or non-detection of these non-Gaussianities provides us a useful insight to find out whether the PBHs can source DM or not.

At the end of this section, it is important to note that in our work, we estimated the abundance of PBHs by assuming the Gaussian statistics for the curvature perturbations. But, as we saw above, our computations imply that this assumption is ruled out around the peak of the scalar power spectrum in this model. On the other hand, the abundance of PBHs is very sensitive to the statistics of the curvature perturbations. Therefore, the estimation of the PBHs abundance in the framework of this model should be reconsidered by using the formulation based on the non-Gaussian statistics, and this is out of the scope of the present paper, and we leave it for future studies.


\section{Secondary gravitational waves}
\label{sec:sgws}

The enhancement of the power spectrum of the curvature perturbations, in addition to the production significant amount of PBHs at special small scales during inflation, may also lead to the generation of secondary GWs. These GWs have gathered considerable attention among the inflationary contexts recently because they can be tested through the data of several earth-based or space-based GW observatories. The detection of GWs can provide crucial information to check the validity of inflationary models. In this section, we examine the secondary GWs in the $\alpha$-attractor inflation scenario with a tiny bump in the potential energy of the inflaton and then compare its predictions with the sensitivity regions of some designed GWs detectors.

Although the spectrum of the secondary GWs depends on the statistics of the primordial density fluctuations, but unlike the PBHs abundance, the sensitivity of the spectrum in this case is not very efficient and it is hardly affected \cite{Zhang:2021vak}. Therefore, it will be useful here to use the formalism which is valid for the Gaussian perturbations to evaluate the induced GWs spectrum to provide some estimations for height and frequency of the spectrum peak in our model. According to this formalism, the fractional energy density of the induced GWs in the radiation dominated era is expressed as \cite{Kohri:2018awv, Lu:2019sti}
\begin{equation}
 \label{OmegaGW}
 \Omega_{GW}(k,\eta)=\frac{1}{6}\left(\frac{k}{aH}\right)^{2}\int_{0}^{\infty}dv\int_{|1-v|}^{|1+v|}du\left(\frac{4v^{2}-\left(1-u^{2}+v^{2}\right)^{2}}{4uv}\right)^{2}\overline{I_{RD}^{2}(u,v,x)}\mathcal{P}_{s}(ku)\mathcal{P}_{s}(kv),
\end{equation}
where $\eta$ indicates the conformal time, and the time average of the source terms is given by
\begin{align}
 \overline{I_{RD}^{2}(u,v,x\to\infty)}= & \frac{1}{2x^{2}}\Bigg[\left(\frac{3\pi\left(u^{2}+v^{2}-3\right)^{2}\Theta\left(u+v-\sqrt{3}\right)}{4u^{3}v^{3}}+\frac{T_{c}(u,v,1)}{9}\right)^{2}
 \nonumber\\
 & +\left(\frac{\tilde{T}_{s}(u,v,1)}{9}\right)^{2}\Bigg].
 \label{IRD2b}
\end{align}
The functions $T_{c}$ and $T_{s}$ in the above equation are defined in the following forms
\begin{align}
T_{c}= & -\frac{27}{8u^{3}v^{3}x^{4}}\Bigg\{-48uvx^{2}\cos\left(\frac{ux}{\sqrt{3}}\right)\cos\left(\frac{vx}{\sqrt{3}}\right)\left(3\sin(x)+x\cos(x)\right)+
\nonumber\\
& 48\sqrt{3}x^{2}\cos(x)\left(v\sin\left(\frac{ux}{\sqrt{3}}\right)\cos\left(\frac{vx}{\sqrt{3}}\right)+u\cos\left(\frac{ux}{\sqrt{3}}\right)\sin\left(\frac{vx}{\sqrt{3}}\right)\right)+
\nonumber\\
& 8\sqrt{3}x\sin(x)\Bigg[v\left(18-x^{2}\left(u^{2}-v^{2}+3\right)\right)\sin\left(\frac{ux}{\sqrt{3}}\right)\cos\left(\frac{vx}{\sqrt{3}}\right)+
\nonumber\\
& u\left(18-x^{2}\left(-u^{2}+v^{2}+3\right)\right)\cos\left(\frac{ux}{\sqrt{3}}\right)\sin\left(\frac{vx}{\sqrt{3}}\right)\Bigg]+
\nonumber\\
& 24x\cos(x)\left(x^{2}\left(-u^{2}-v^{2}+3\right)-6\right)\sin\left(\frac{ux}{\sqrt{3}}\right)\sin\left(\frac{vx}{\sqrt{3}}\right)+
\nonumber\\
& 24\sin(x)\left(x^{2}\left(u^{2}+v^{2}+3\right)-18\right)\sin\left(\frac{ux}{\sqrt{3}}\right)\sin\left(\frac{vx}{\sqrt{3}}\right)\Bigg\}
\nonumber\\
& -\frac{\left(27\left(u^{2}+v^{2}-3\right)^{2}\right)}{4u^{3}v^{3}}\Bigg\{\text{Si}\left[\left(\frac{u-v}{\sqrt{3}}+1\right)x\right]-\text{Si}\left[\left(\frac{u+v}{\sqrt{3}}+1\right)x\right]
\nonumber\\
& +\text{Si}\left[\left(1-\frac{u-v}{\sqrt{3}}\right)x\right]-\text{Si}\left[\left(1-\frac{u+v}{\sqrt{3}}\right)x\right]\Bigg\},
\label{Tc}
\end{align}

\begin{align}
T_{s}= & \frac{27}{8u^{3}v^{3}x^{4}}\Bigg\{48uvx^{2}\cos\left(\frac{ux}{\sqrt{3}}\right)\cos\left(\frac{vx}{\sqrt{3}}\right)\left(x\sin(x)-3\cos(x)\right)-
\nonumber\\
& 48\sqrt{3}x^{2}\sin(x)\left(v\sin\left(\frac{ux}{\sqrt{3}}\right)\cos\left(\frac{vx}{\sqrt{3}}\right)+u\cos\left(\frac{ux}{\sqrt{3}}\right)\sin\left(\frac{vx}{\sqrt{3}}\right)\right)+
\nonumber\\
& 8\sqrt{3}x\cos(x)\Bigg[v\left(18-x^{2}\left(u^{2}-v^{2}+3\right)\right)\sin\left(\frac{ux}{\sqrt{3}}\right)\cos\left(\frac{vx}{\sqrt{3}}\right)+
\nonumber\\
& u\left(18-x^{2}\left(-u^{2}+v^{2}+3\right)\right)\cos\left(\frac{ux}{\sqrt{3}}\right)\sin\left(\frac{vx}{\sqrt{3}}\right)\Bigg]+
\nonumber\\
& 24x\sin(x)\left(6-x^{2}\left(-u^{2}-v^{2}+3\right)\right)\sin\left(\frac{ux}{\sqrt{3}}\right)\sin\left(\frac{vx}{\sqrt{3}}\right)+
\nonumber\\
& 24\cos(x)\left(x^{2}\left(u^{2}+v^{2}+3\right)-18\right)\sin\left(\frac{ux}{\sqrt{3}}\right)\sin\left(\frac{vx}{\sqrt{3}}\right)\Bigg\}-\frac{27\left(u^{2}+v^{2}-3\right)}{u^{2}v^{2}}+
\nonumber\\
& \frac{\left(27\left(u^{2}+v^{2}-3\right)^{2}\right)}{4u^{3}v^{3}}\Bigg\{-\text{Ci}\left[\left|1-\frac{u+v}{\sqrt{3}}\right|x\right]+\ln\left|\frac{3-(u+v)^{2}}{3-(u-v)^{2}}\right|+
\nonumber\\
& \text{Ci}\left[\left(\frac{u-v}{\sqrt{3}}+1\right)x\right]-\text{Ci}\left[\left(\frac{u+v}{\sqrt{3}}+1\right)x\right]+\text{Ci}\left[\left(1-\frac{u-v}{\sqrt{3}}\right)x\right]\Bigg\}.
\label{Ts}
\end{align}
In these equations, the sine-integral $\text{Si}(x)$ and cosine-integral $\text{Ci}(x)$ functions are defined respectively as follows
\begin{equation}
 \label{SiCi}
 \text{Si}(x)=\int_{0}^{x}\frac{\sin(y)}{y}dy,\qquad\text{Ci}(x)=-\int_{x}^{\infty}\frac{\cos(y)}{y}dy.
\end{equation}
The relation between the current energy densities of the included GWs and the corresponding values well after their horizon reentry in the radiation domination era, is given by the following equation
\begin{equation}
 \label{OmegaGW0}
 \Omega_{GW}\left(k,\eta_{0}\right)=\Omega_{GW}(k,\eta)\frac{\Omega_{r0}}{\Omega_{r}(\eta)},
\end{equation}
where $\Omega_r$ denotes the fractional energy density of radiation, and the subscript $0$ refers to the present epoch. Here, we take $\Omega_{r0}h^{2}\simeq4.2\times10^{-5}$ \cite{Cai:2019bmk, Fu:2019vqc, Fu:2020lob}. It is worth mentioning that the conformal time $\eta\gg\eta_{k}$ in Eq. \eqref{OmegaGW0} is considered to be earlier than the matter-radiation equality, and of course late enough so that $\Omega_{GW}(k,\eta)$ be converged to a constant value during the radiation dominated era.

\begin{figure}
\begin{center}
\scalebox{0.8}[0.8]{\includegraphics{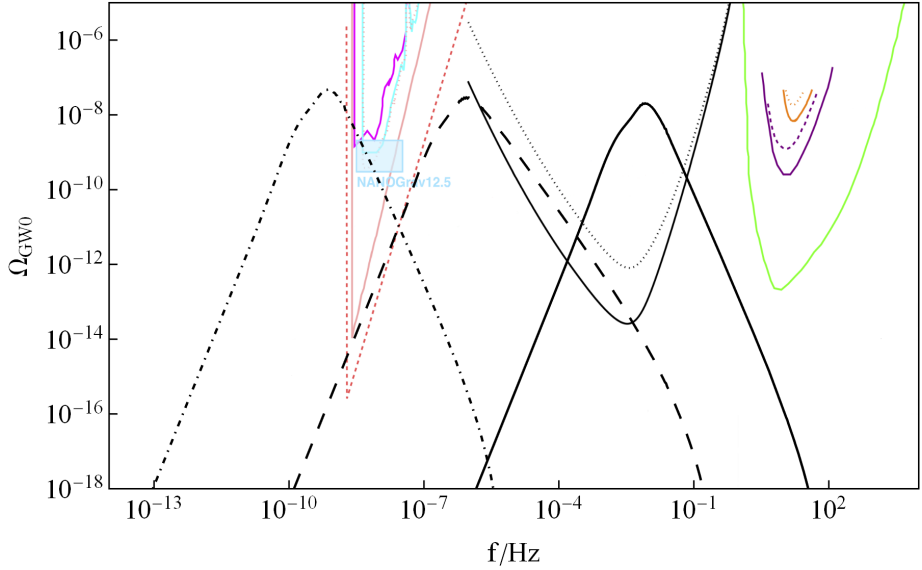}}
\caption{The present fractional energy density of the secondary GWs in terms of frequency. The thick solid, thick dashed, and thick dash-dotted black plots correspond to Cases 1, 2, and 3, respectively. The sensitivity regions of the GWs ground-based interferometers including aLIGO O2 (dotted orange) and O3 (solid orange) runs \cite{LIGOScientific:2017zlf, LIGOScientific:2017vwq,KAGRA:2021kbb}, projections for the HLV (LIGO-Hanford, LIGOLivingston and Virgo) network at design sensitivity (dashed purple), the mid-scale upgrade of aLIGO (A+ detectors, solid purple) and the ET (light green) \cite{Punturo:2010zz} are also demonstrated in the figure \cite{Bagui:2021dqi}. Moreover, the sensitivity of the future space-based interferometer LISA is shown (best experimental design (solid black) and worst experimental design (dashed black)) \cite{LISA:2017pwj}. The Pulsar Timing Arrays (PTA’s) considered here are the EPTA (solid pink) \cite{lentati2015european}, North American Nanohertz Observatory for Gravitational Waves (NANOGrav) (solid light blue) \cite{NANOGrav:2020gpb}, International Pulsar Timing Array (IPTA) (solid light pink), and the Square Kilometer Array (SKA) (dashed red) \cite{Moore:2014lga,Janssen:2014dka,Schmitz:2020syl}.}
\label{fig:OmegaGW0}
\end{center}
\end{figure}

In Fig. \ref{fig:OmegaGW0}, the fractional energy density of the secondary GWs is plotted versus frequency for the three sets of parameters that are listed in Table \ref{table:parameters}, by using Eq. \eqref{OmegaGW0}. The sensitivity regions of the GWs ground-based interferometers including aLIGO O2 (dotted orange) and O3 (solid orange) runs \cite{LIGOScientific:2017zlf,LIGOScientific:2017vwq,KAGRA:2021kbb}, projections for the HLV (LIGO-Hanford, LIGOLivingston and Virgo) network at design sensitivity (dashed purple), the mid-scale upgrade of aLIGO (A+ detectors, solid purple) and the ET (light green) \cite{Punturo:2010zz} are also marginalized in the figure \cite{Bagui:2021dqi}. Moreover, the sensitivity of the future space-based interferometer LISA is shown (best experimental design (solid black) and worst experimental design (dashed black)) \cite{LISA:2017pwj}. The Pulsar Timing Arrays (PTA’s) considered here are the EPTA (solid pink) \cite{lentati2015european}, North American Nanohertz Observatory for Gravitational Waves (NANOGrav) (solid light blue) \cite{NANOGrav:2020gpb}, International Pulsar Timing Array (IPTA) (solid light pink), and the Square Kilometer Array (SKA) (dashed red) \cite{Moore:2014lga,Janssen:2014dka,Schmitz:2020syl}. Fig. \ref{fig:OmegaGW0} clearly illustrates that the peak amplitude of the spectra is of order $10^{-8}$ for all three cases, but the peaks occur in different frequencies. For Case 1, the peak appears at the critical frequency $f_{c}\sim {\cal O}(10^{-3})\,\mathrm{Hz}$, and the result is located within the joint region of LISA (best and worst) detector. The peak of Case 2 appears at the critical frequency $f_{c}\sim {\cal O} (10^{-7})\,\mathrm{Hz}$, and the result can be located within the sensitivity regions of LISA (best), and SKA. For Case 3, the peak takes place at the frequency $f_{c}\sim {\cal O}(10^{-10})\,\mathrm{Hz}$, and it can lie within the sensitivity regions of SKA, IPTA, and EPTA. The spectrum of Case 3 can also be located within the region of the detected NANOGrav signal, and therefore the induced GWs in our PBHs formation scenario are capable to explain this observation. We present the exact values of the critical frequencies and peak heights for the three cases in Table \ref{table:GWs}. As we see, the predictions of our model in three cases, lie inside the sensitivity marginalized joints regions of some designed GWs detectors. It is a remarkable consequence of our scenario that may be regarded in the future to check the viability of the model in light of the observational data.

Here, we also estimate the tilt of the GWs spectrum at different ranges of frequency. Recently, it has been argued that the power spectrum of $\Omega_{\rm GW0}$ can be expressed in terms of frequency as the power-law form $\Omega_{\rm GW0}\sim f^{n}$, where $n$ is a constant parameter \cite{Xu:2019bdp, Fu:2019vqc, Kuroyanagi:2018csn}. In the setup of $\alpha$-attractor inflation with a tiny bump, we compute the parameter $n$ for the frequency bands $f \ll f_c$, $f < f_c$, and $f > f_c$. We denote the results of these regions by $n_{f\ll f_{c}}$, $n_{f<f_{c}}$, and $n_{f>f_{c}}$, respectively. Our numerical findings are presented in Table \ref{table:GWs}. The results obtained in the infrared regime $f \ll f_c$ verify properly the analytical relation $\Omega_{\rm GW0}\sim f^{3-2/\ln\left(f_{c}/f\right)}$ derived by \cite{Yuan:2019wwo, Cai:2019cdl}.

\begin{table}[ht!]
  \centering
  \caption{The frequencies and heights of the peak of the spectrum of the current fractional energy density of the secondary GWs for Cases 1, 2, and 3.}
\scalebox{1}[1] {
\begin{tabular}{cccccc}
\hline
\#  & $\qquad\qquad$ $f_{c}$ $\qquad\qquad$ & $\quad$ $\Omega_{\rm GW0}\left(f_{c}\right)$ $\quad$ & $\quad$ $n_{f\ll f_{c}}$ $\quad$ & $\quad$ $n_{f<f_{c}}$ $\quad$ & $\quad$ $n_{f>f_{c}}$\tabularnewline
\hline
\hline
Case 1 & $8.163\times10^{-3}$ & $2.081\times10^{-8}$ & $3.010$ & $1.584$ & $-2.841$ \tabularnewline
\hline
Case 2 & $8.807\times10^{-7}$ & $2.954\times10^{-8}$ & $3.021$ & $1.582$ & $-1.702$\tabularnewline
\hline
Case 3 & $7.389\times10^{-10}$ & $4.832\times10^{-8}$ & $3.016$ & $1.559$ & $-2.736$\tabularnewline
\hline
\end{tabular}
    }
  \label{table:GWs}
\end{table}


\section{Conclusions}
\label{section:conclusions}

We studied the process of PBHs formation in the setup of $\alpha$-attractor inflation with a tiny bump in its potential \cite{mishra2020primordial}. For this purpose, we focused on three cases which the values of these cases are listed in Table \ref{table:parameters}. We evaluated the scalar power spectrum in terms of the comoving wavenumber by solving numerically the Mukhanov-Sasaki equation for these cases and showed that in each case, the power spectrum exhibits a peak on a special scale that is beyond the CMB scales. The diagrams of the scalar power spectrum are shown in Fig. \ref{fig:Ps}. The height of the peak in all cases is of order $10^{-2}$, which is seven orders of magnitudes larger than the one observed in the CMB scales ($\mathcal{P}_{s*} \approx 2.1 \times 10^{-9}$) \cite{akrami2020planck}. This enhancement is large enough for the formation of PBHs. Therefore, we confirmed that the inclusion of a tiny bump in the $\alpha$-attractor potential can explain successfully the PBHs production from inflation. The abundance of PBHs for each case in this setting has also been estimated, and it has been shown that in this model the PBHs can be formed in a wide range of mass scales. The diagrams of the abundance of the produced PBHs in terms of the mass are plotted in Fig. \ref{fPBH-M}. We concluded that the formed PBHs in Case 1, can constitute a significant contribution to the DM observed today. We compared the results of our model for the curvature power spectrum with the current observational bounds on this observable, and found that all the three cases satisfy the constraints properly. The results of these cases are also compatible with the observational bounds on abundance of PBHs.

Subsequently, we turned to calculate the inflationary non-Gaussianity for the given three cases. For this purpose, we applied the computational code BINGO \cite{Hazra:2012yn} which calculates the primordial non-Gaussianity in different triangle configurations. At the CMB scales, the results of the models for different types of non-Gaussianity are of the order of the inflationary slow-roll parameter, $f_{\mathrm{NL}}^{\mathrm{equil}}\sim\mathcal{O}\left(10^{-2}\right)$, and therefore they are in good agreement with the Planck 2018 constraints \cite{Planck:2019kim}. However, away from the CMB scales the equilateral non-Gaussianity in this model gets amplified and can obtain values much larger than unity which are of order $f_{\mathrm{NL}}^{\mathrm{equil}}\sim\mathcal{O}\left(10^{3}\right)$. The peaks of the equilateral non-Gaussianity for the considered cases appear at the scales which are related to the BBN \cite{Jeong:2014gna, Inomata:2016uip}, $\mu$-distortion \cite{Fixsen:1996nj} events or smaller scales. We also found out that the equilateral non-Gaussianity for each case peaks at a scale close to the one for the peak of the power spectrum of the scalar perturbations. In fact, the peaks of the equilateral non-Gaussianity take place at the scales where the scalar power spectrum reveals a sudden declination in its diagram versus comoving wavenumber before it gets amplified and reaches its maximum value. This result is in agreement with the findings of \cite{Davies:2021loj} for an ultra-slow-roll inflation model, and with the findings of \cite{Zhang:2021rqs} in the setup of Gauss-Bonnet inflation. The existence of such a sharp declination appears also in some other models like constant-roll \cite{Motohashi:2019rhu} and ultra constant-roll \cite{Liu:2020oqe} inflation models. Due to this property, we expect that these models result in a substantial enhancement for the equilateral non-Gaussianity too, and this prediction can be checked by further investigations. Contrariwise, there are some models, such as the non-minimal derivative \cite{Fu:2019ttf, Dalianis:2019vit, Teimoori:2021thk} and $k/G$ \cite{Lin:2020goi, Yi:2020cut, Gao:2020tsa, Gao:2021vxb, Solbi:2021rse, Solbi:2021wbo, Teimoori:2021pte, Heydari:2021gea, Heydari:2021qsr} inflationary scenarios, where the variation of the scalar power spectrum is rather mild in all scales, and thus we expect that these models do not show a large enhancement in the equilateral non-Gaussianity and it remains of order $f_{\mathrm{NL}}^{\mathrm{equil}}\lesssim1$ in all scales. This point also may be verified by future examinations.

We then used the BINGO code to compute the non-Gaussianity in the general form of a scalene triangle. The results of this configuration for the considered three cases are presented in two-dimensional diagrams of Figs. \ref{fig:scalene-case1}, \ref{fig:scalene-case2}, and \ref{fig:scalene-case3}. From these diagrams, we concluded that in the $\alpha$-attractor inflation with a tiny bump in the potential, the non-Gaussianity peaks in the equilateral triangle configuration, and therefore the equilateral non-Gaussianity is the most prominent type of non-Gaussianity in this scenario. Our investigation implies that the $\alpha$-attractor model of inflation with a tiny bump in its potential predicts significant non-Gaussianities at scales beyond the CMB scales which include the BBN \cite{Jeong:2014gna, Inomata:2016uip} and $\mu$-distortion \cite{Fixsen:1996nj} scales or smaller scales. If such non-Gaussianities are observed in the future measurements, then our model will be confirmed or will be constrained. Moreover, observation or non-observation of such non-Gaussianities provides a powerful insight for us to understand whether the PBHs from inflation can be regarded as the origin of DM or not. We further conclude that study of non-Gaussianity can discriminate between the models proposed so far for PBHs production through inflation.

It is important to note that in our investigation, we estimated the abundance of PBHs for the $\alpha$-attractor scenario with a tiny bump by assuming Gaussian statistics for the curvature perturbations. However, as we discussed so far, this scenario manifests strong non-Gaussianities around the peak of the scalar power spectrum, and on the other hand, the abundance of PBHs is very sensitive to the statistical properties of the primordial curvature fluctuations. Therefore, the estimation of the PBHs abundance in this scenario should be reconsidered by applying the formalism which is valid for the non-Gaussian perturbations. We leave this issue for further investigations in the future.

We further studied the secondary GWs induced in the process of PBHs production in our setup. Although the spectrum of the included GWs also depends on the statistics of the scalar perturbations, but in this case, unlike the PBHs abundance, the sensitivity is not high such that the results of the Gaussian and non-Gaussian treatments are almost close to each other with a good approximation \cite{Zhang:2021vak}. To study the secondary GWs in our scenario, we followed the same procedure which is based on the Gaussian perturbations to provide some assessments for the height and frequency of the spectrum peak for the different cases of our model. In particular, we computed the present fractional energy density ($\Omega_{\rm GW0}$) for the three sets of parameters listed in Table \ref{table:parameters}. We showed that for all the three cases, the spectra of $\Omega_{\rm GW0}$ represent their peaks at different frequencies, while the height of the peaks is of order $10^{-8}$ for all the cases. For Case 1, the frequency of the peak is $8.163\times10^{-3} \mathrm{Hz}$, and it is located inside the sensitivity region of LISA (best and worst) detector. For cases 2 and 3, the frequencies of the peaks appeared at $8.807\times10^{-7} \mathrm{Hz}$ and $7.389\times10^{-10}\,\mathrm{Hz}$, respectively. The peak of the spectrum of $\Omega_{\rm GW0}$ for case 2 can lie within the sensitivity regions of LISA (best), and SKA, and for case 3 within the sensitivity regions of SKA, IPTA, and EPTA. It is interesting that the spectrum of Case 3 can enter the observational joint region of the NANOGrav signal, and therefore the induced GWs in our PBHs formation scenario are capable to explain this signal. Since the results of our scenario can be located inside the detectable limits of GWs detectors, it seems that we can hope to test our model in light of the future observational data.

Additionally, we presented some estimations for the tilt of the secondary GWs spectrum in our scenario for different frequency bands. Our results imply that the power spectrum of induced GWs behaves almost as the power-law function $\Omega_{\rm GW0}\sim f^{n}$ in various intervals of frequency. For each case of our model, we presented the values of the constant parameter $n$ for different frequency domains and showed that the results for the infrared regime $f \ll f_c$ are in good agreement with the analytical expression $\Omega_{\rm GW0}\sim f^{3-2/\ln\left(f_{c}/f\right)}$ obtained by \cite{Yuan:2019wwo, Cai:2019cdl}.

\end{document}